%% file: main.tex
\documentclass[times,twocolumn,final]{elsarticle}

\usepackage{framed,multirow}

\usepackage{mathtools}
\usepackage{amsmath}
\usepackage{array,varwidth}
\usepackage[para,online,flushleft]{threeparttable}
\usepackage{xcolor}

\usepackage{amssymb}
\usepackage{latexsym}
\usepackage{comment}
\usepackage[switch]{lineno}
\usepackage{gensymb}

\newcommand\RotatedNameEntry[1]{%
\parbox[t]{2mm}{\multirow{12}{*}{\rotatebox[origin=c]{90}{Misalignment level in UKB data}}}}

\newcommand\NameEntry[1]{%
  \multirow{3}*{%
    \begin{varwidth}{2cm}
    \flushright #1%
    \end{varwidth}}}

\makeatletter
\def\Cline#1#2{\@Cline#1#2\@nil}
\def\@Cline#1-#2#3\@nil{%
  \omit
  \@multicnt#1%
  \advance\@multispan\m@ne
  \ifnum\@multicnt=\@ne\@firstofone{&\omit}\fi
  \@multicnt#2%
  \advance\@multicnt-#1%
  \advance\@multispan\@ne
  \leaders\hrule\@height#3\hfill
  \cr}
\makeatother

\usepackage{url}
\usepackage{xcolor}

\usepackage{hyperref}

\definecolor{newcolor}{rgb}{.8,.349,.1}

\begin{document}


\begin{frontmatter}

\title{Multi-class point cloud completion networks for 3D cardiac anatomy reconstruction from cine magnetic resonance images}%

\input{abstract-authors}

\end{frontmatter}


\section{Introduction}
\label{sec:introduction}

Cardiac magnetic resonance imaging (MRI) is the gold standard for the assessment of a large number of cardiovascular pathologies due its excellent soft-tissue contrast, lack of ionizing radiation, and minimal use of contrast agents \citep{stokes2017role}. In current clinical practice, most cine cardiac MRI acquisitions consist of a stack of two-dimensional (2D) short-axis (SAX) slices that provide a cross-sectional view of the heart, as well as multiple 2D long-axis (LAX) slices that intersect the heart longitudinally at different angles. While this allows the visualization of cardiac anatomy from multiple different views, the cine slices only capture information in 2D planes and are therefore unable to truly represent the inherent three-dimensional (3D) structure of the heart \citep{o2019accuracy}. However, accurate 3D cardiac anatomy models are necessary for a wide variety of applications in both clinical practice and research settings, including the accurate measurement of image-based biomarkers, discovery of novel biomarkers, visualization of healthy and pathological cardiac anatomy, and the development of both population-wide and case-specific modelling of cardiac mechanics and electrophysiology \citep{yang20173d,gilbert2019independent,attar20193d,minchole2019mri,mauger2019right,corral2020digital,levrero2020sensitivity,beetz2021predicting,beetz2022multi,beetz2022combined,beetz2022interpretable,mauger2022multi,beetz2023mesh,beetz2023post}.

Consequently, multiple research efforts have been dedicated to developing MRI-based methods capable of creating 3D representations of the human heart. The first group of approaches attempts to achieve this by increasing the spatial resolution of the MRI acquisition itself, \emph{e.g.} 3D MRI \citep{mascarenhas2006fast,jeong2015single}. However, most techniques suffer from lower temporal resolution and reduced image quality compared to 2D acquisitions, making an accurate assessment of cardiac function more difficult \citep{amano2017three,usman2017free}. While more recent works have improved these shortcomings considerably, they were only tested on a small number of cases, are dependent on the availability of the most recent scanner hardware and software, come with long reconstruction times, and often only allow partial heart coverage \citep{wetzl2018single,kustner2020isotropic,kustner2020cinenet}.

A second group of approaches aims at reconstructing true 3D representations of the heart from the available clinically standard 2D cine MRI slices, which is also the focus of this work. These techniques typically first segment the cardiac structures of interest in the images and then use the resulting contours to reconstruct the corresponding 3D anatomical surface models. 
Similar to many other medical image analysis tasks, deep learning methods, such as the fully convolutional neural network (FCN) \citep{long2015fully}, U-Net \citep{ronneberger2015u}, and their derivatives \citep{cciccek20163d}, have become the state-of-the-art approach for cardiac image segmentation in the recent past \citep{chen2020deep}. While most cardiac MRI segmentation research has focused on biventricular segmentation from SAX images \citep{chen2020deep}, some have also included the two-chamber (2ch LAX) and four-chamber long-axis (4ch LAX) views \citep{bai2018automated}. More recent efforts aim to extend this success to more complex, multi-domain imaging datasets suffering from domain shift \citep{campello2021multi,eisenmann2022biomedical,martin2023deep}. Hereby, the task is to either detect erroneous segmentations with improved quality control measures \citep{tarroni2020large,wang2020deep,machado2021quality} or avoid such failures altogether by using more robust algorithms that incorporate topological information about the underlying anatomy \citep{oktay2017anatomically,byrne2020persistent}. 

Given a set of contours derived from the 2D segmentation masks, the task of 3D surface reconstruction is a challenging, ill-posed optimization problem for two main reasons. First, the available 2D information is extremely sparse as compared to a 3D representation making an accurate surface reconstruction difficult, especially in regions with little or no data. In addition, artifacts caused by various types of motion (cardiac, respiratory, patient) during image acquisition result in slice misalignment and potentially erroneous anatomical information \citep{sievers2005impact,scott2009motion,bogaert2012clinical}.

Some research has focused on developing methods specifically for 2D misalignment correction in order to facilitate the following 3D reconstruction task. These include slice-to-slice registration \citep{goshtasby1996fusion,mcleish2002study,villard2016correction}, slice-to-volume-registration \citep{chandler2008correction,su2014automatic,ferrante2017slice}, probabilistic segmentation maps generated with decision forests \citep{tarroni2018comprehensive}, combined image slice segmentation and alignment correction \citep{villard2018isachi}, and statistical shape model based misalignment correction \citep{banerjee2021miua}. Considerable research efforts have also focused on directly addressing these challenges as part of the 3D surface reconstruction task \citep{villard2018surface,mauger2019right,banerjee2021ptrsa}. In recent times, both grid-based \citep{xu2019ventricle} and geometric deep learning methods \citep{beetz2021biventricular,chen2021shape,beetz2022reconstructing,beetz2023point2mesh} have been increasingly used for cardiac surface reconstruction from various types of inputs, such as single 2D images \citep{zhou2019one,wang2020instantiation}, the SAX stack \citep{chen2021shape}, or both SAX and LAX images \citep{xu2019ventricle,beetz2021biventricular,beetz2022reconstructing,beetz2023point2mesh}.

In this work, we propose to utilize recent advances in geometric deep learning on point clouds \citep{qi2017pointnet,qi2017pointnet++} to design a novel cardiac surface reconstruction method. Of particular importance for this work is 3D point cloud completion, which tries to predict the complete shape of a point cloud surface from a partial input \citep{yang2017foldingnet,achlioptas2018learning,yuan2018pcn}. Point cloud based deep learning methods have recently also been applied to various cardiac image analysis tasks, including segmentation \citep{ye2020pc}, anatomy generation \citep{beetz2021gen}, deformation prediction \citep{beetz2021predicting}, pathology classification \citep{chang2020automatic,beetz20233d}, and the combined modeling of cardiac anatomy and electrophysiology data \citep{beetz2022multi,beetz2022combined,li2022deep}.

To the best of our knowledge, this work is the first point cloud-based deep learning approach for multi-class bitemporal cardiac anatomy reconstruction from 2D cine MRI slices. Previous approaches lacked validation on real data and used inefficient voxel grid representations \citep{xu2019ventricle}, did not incorporate class-specific and temporal information \citep{beetz2021biventricular}, or relied on the different approach of mesh template deformation with graph neural networks while only using SAX information \citep{chen2021shape}. Our main contributions are summarized as follows:

\begin{itemize}
	\item We develop a 3D biventricular surface reconstruction pipeline with a novel point cloud-based deep learning network capable of addressing the data sparsity, motion artifact, and potential errors introduced as part of the segmentation or contouring process into a single model, while at the same time maintaining both multi-class and bitemporal anatomy information;
	\item We evaluate our proposed multi-class point cloud completion network (PCCN) on a large-scale dataset of synthetic biventricular anatomies and demonstrate highly accurate reconstruction performance in a multi-temporal setting, at both the diastolic and systolic ends of the cardiac cycle;
	\item We compare our PCCN to a state-of-the-art 3D U-Net approach and show its advantages in terms of reconstruction results and efficiency in data representation;
	\item We successfully apply and validate the complete reconstruction pipeline on cine MRI acquisitions of 1000 UK Biobank (UKB) cases;
	\item We calculate common clinical metrics from our method's UKB reconstructions and find plausible values compared to other population-wide cardiac anatomy studies; and
	\item We conduct a robustness analysis of our PCCN with respect to erroneous input contours and increasing levels of misalignment.
\end{itemize}

The rest of the paper is organized as follows:
Sec.~\ref{sec:dataset} describes the two datasets used for method development and evaluation in this work. A detailed description of our proposed pipeline is provided in Sec.~\ref{sec:methods}, while the experiments conducted for method evaluation along with the corresponding results are presented in Sec.~\ref{sec:experiments}. Finally, Sec.~\ref{sec:discussion} provides a discussion of the proposed technique and our experimental findings, before Sec.~\ref{sec:conclusion} concludes the paper.

\section{Datasets}
\label{sec:dataset}

We use both a synthetic dataset generated from a high-resolution statistical shape model (SSM) (Sec.~\ref{sec:statistical_shape_model}) and the real cine MRI acquisitions of the UK Biobank study (Sec.~\ref{sec:uk_biobank_dataset}) to develop and evaluate our method.

\subsection{3D MRI-based statistical shape model}
\label{sec:statistical_shape_model}

The first dataset of this work is based on the biventricular shape model from \citet{bai2015bi}, which was created from the 3D cardiac MRI scans of 1084 healthy volunteers with a 3D cine balanced steady-state free precession (b-SSFP) sequence and a resolution of $1.25 \times 1.25 \times 2$~mm. The authors registered and segmented all images at the end-diastolic (ED) and end-systolic (ES) phases of the cardiac cycle to construct two 3D biventricular surface meshes and applied principal component analysis to determine the 100 most important modes of variation of these two mean shapes. We use this SSM to derive a population of 3D biventricular anatomies and corresponding sparse 2D cine MRI inputs to train and evaluate our PCCN (Sec.~\ref{sec:dataset_generation}). The SSM was selected as a basis for our synthetic data generation process due to multiple reasons. First, it is based on 3D MRI acquisitions, which offer high spatial resolution both in-plane and between image planes without the effects of slice misalignment and data sparsity. Second, the dataset was derived from a large and representative number of volunteers, increasing its robustness and ability to accurately capture the true variability in the population. Third, only healthy individuals were considered and consistent scanning protocols were used, making it compatible with large-scale cardiac imaging studies such as the UK Biobank dataset. Hence, we consider the shapes generated from the SSM as the ground truth for our method development.

\subsection{UK Biobank}
\label{sec:uk_biobank_dataset}

The second dataset used in this work consists of the 2D cine MRI acquisitions of 500 male and 500 female cases randomly selected from the UK Biobank study \citep{petersen2013imaging,petersen2015uk}. For each case, we consider the first temporal frame of the cine sequence as the ED phase of the cardiac cycle and determine the frame of the ES phase from the segmented SAX stack as the cardiac phase with minimum LV volume \citep{banerjee2021ptrsa}. As our dataset, we select all SAX slices as well as the two-chamber (2ch) LAX and four-chamber (4ch) LAX slices for both ED and ES phases of the cine sequence for each case. Its large sample size and typical image resolution ($1.8 \times 1.8 \times 8.0$~mm), the availability of metadata for each case (sex, age), and the usage of a clinically established acquisition protocol (b-SSFP) make it an ideal choice for the evaluation of our proposed cardiac surface reconstruction pipeline under real-world conditions. Including both ED and ES phases in the dataset allows us to additionally analyze the performance of our pipeline in a multi-temporal setting, which is crucial for many follow-up cardiac function tasks \citep{beetz2021predicting,beetz2022combined}.

\begin{figure*}[!ht]
	\centerline{\includegraphics[width=0.9\textwidth]{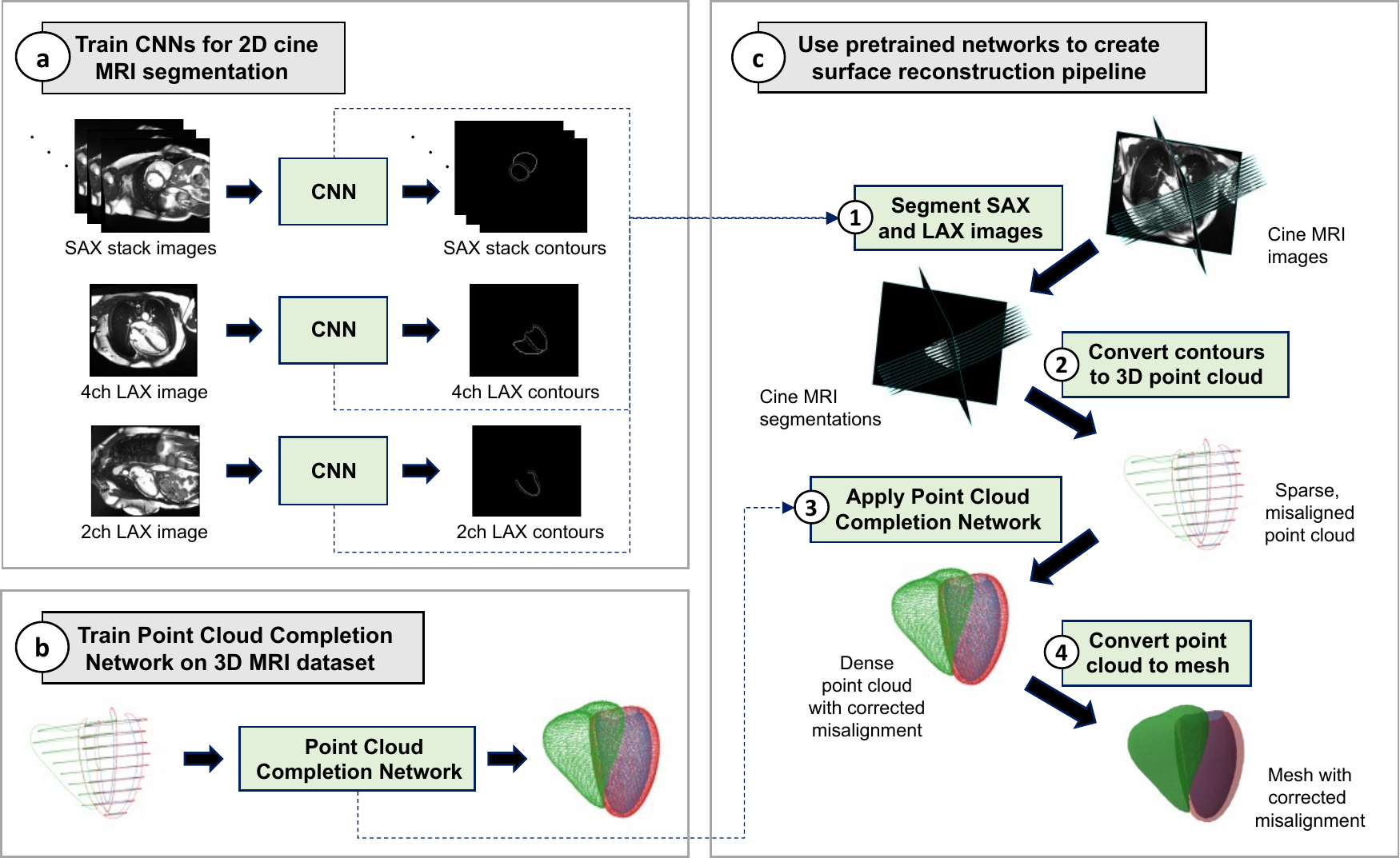}}
	\caption{Overview of our proposed 3D cardiac surface reconstruction pipeline from cine MR images. (a) We train three separate CNNs to segment SAX, 4ch LAX, and 2ch LAX cine MR images. (b) We train a Point Cloud Completion Network on 3D MRI-based dataset to reconstruct a dense 3D point cloud with corrected misalignment from a sparse, misaligned input point cloud. (c) We propose a 4-step pipeline to reconstruct 3D multi-class cardiac meshes from raw cine MRI acquisitions using the pre-trained networks (a,b) in steps 1 and 3 of the pipeline.}
	\label{fig:reconstruction_pipeline}
\end{figure*}

\section{Methods}
\label{sec:methods}

In this work, we propose a fully-automatic 3D biventricular surface reconstruction pipeline consisting of four steps outlined in Fig.~\ref{fig:reconstruction_pipeline}-c. First, three pre-trained convolutional neural networks (CNN) are applied to segment the SAX, 4ch LAX, and 2ch LAX slices of the input cine MRI acquisition (Sec.~\ref{sec:mri_segmentation}). Second, the anatomical contours obtained from the segmentation step are positioned in 3D space and converted into point clouds (Sec.~\ref{sec:contours_to_point_cloud}). Third, a pre-trained Point Cloud Completion Network (PCCN) is used to reconstruct a dense multi-class point cloud representation of the biventricular anatomy from the sparse, misaligned input point cloud in what constitutes the key step of the pipeline (Sec.~\ref{sec:point_completion_network}). Finally, the dense point cloud is transformed into an anatomical mesh (Sec.~\ref{sec:meshing_step}. The pre-training step of both the CNNs (Fig.~\ref{fig:reconstruction_pipeline}-a) and PCCN (Fig.~\ref{fig:reconstruction_pipeline}-b) is conducted before the application of the full reconstruction pipeline. The following subsections describe the four steps of the pipeline in greater detail.

\subsection{Cine MRI segmentation}
\label{sec:mri_segmentation}

The first step consists of the segmentation of the SAX, 4ch LAX, and 2ch LAX image slices of the cine MRI acquisitions of the UK Biobank dataset. To this end, we employ the fully convolutional network (FCN)-based approach proposed by \citet{bai2018automated} for the segmentation of the SAX stack and 4ch LAX slices, since it has been shown to segment heart structures from cine MR slices with human-level accuracy. A detailed description of the segmentation method is provided in the Supplementary Material.

Due to the lack of a publicly available pre-trained network for automated UK Biobank 2ch LAX slice segmentation, we also train a separate conditional generative adversarial network \citep{isola2017image} with a U-Net generator \citep{rezaei2017conditional} for this task. Hereby, the training data consists of 200 2ch LAX frames chosen at random from separate UK Biobank subjects, equally distributed across the whole cardiac sequence. We extract endocardial and epicardial contours, as well as valvular contours along the mitral valve using the open source tool ImageJ \citep{schneider2012nih,rueden2017imagej2} from which segmentation masks are computed. Image and segmentation mask pairs are rigidly augmented using rotations, translations, and crops around the LV center, yielding 1500 and 250 training and validation pairs respectively.

\subsection{Conversion of 2D contours to 3D point cloud}
\label{sec:contours_to_point_cloud}

The objective of the second step of our reconstruction pipeline is to convert the 2D segmentation masks of the different views (SAX stack, 4ch LAX, 2ch LAX) obtained in the previous step into a 3D sparse representation of the cardiac anatomy. To this end, we first extract the LV endocardial, LV epicardial, and RV endocardial contours from their respective segmentation masks. We then fit a B-spline curve separately to each contour and resample the same number of points as in the original contour along the obtained curve at equidistant intervals. We repeat this procedure for both SAX and LAX images and finally place all resulting points in the same 3D space as the original cine MR slices to create the corresponding biventricular point clouds for each case.

\subsection{Multi-class point cloud completion network}
\label{sec:point_completion_network}

The third step of our pipeline aims to address both the sparsity and misalignment challenges of cardiac surface reconstruction with a single deep learning model, while maintaining the spatial and temporal information of all anatomical structures. To this end, we propose a novel multi-class Point Cloud Completion Network, which acts directly on the sparse, misaligned point cloud representations of the biventricular anatomy.
The following subsections explain the network architecture (Sec.~\ref{sec:network_architecture}, loss function and training procedure of the PCCN (Sec.~\ref{sec:loss_function}, including the generation process of a synthetic biventricular anatomy dataset (Sec.~\ref{sec:dataset_generation} for network training and an initial validation.

\subsubsection{Network architecture}
\label{sec:network_architecture}

\begin{figure}[!t]
	\centerline{\includegraphics[width=0.485\textwidth]{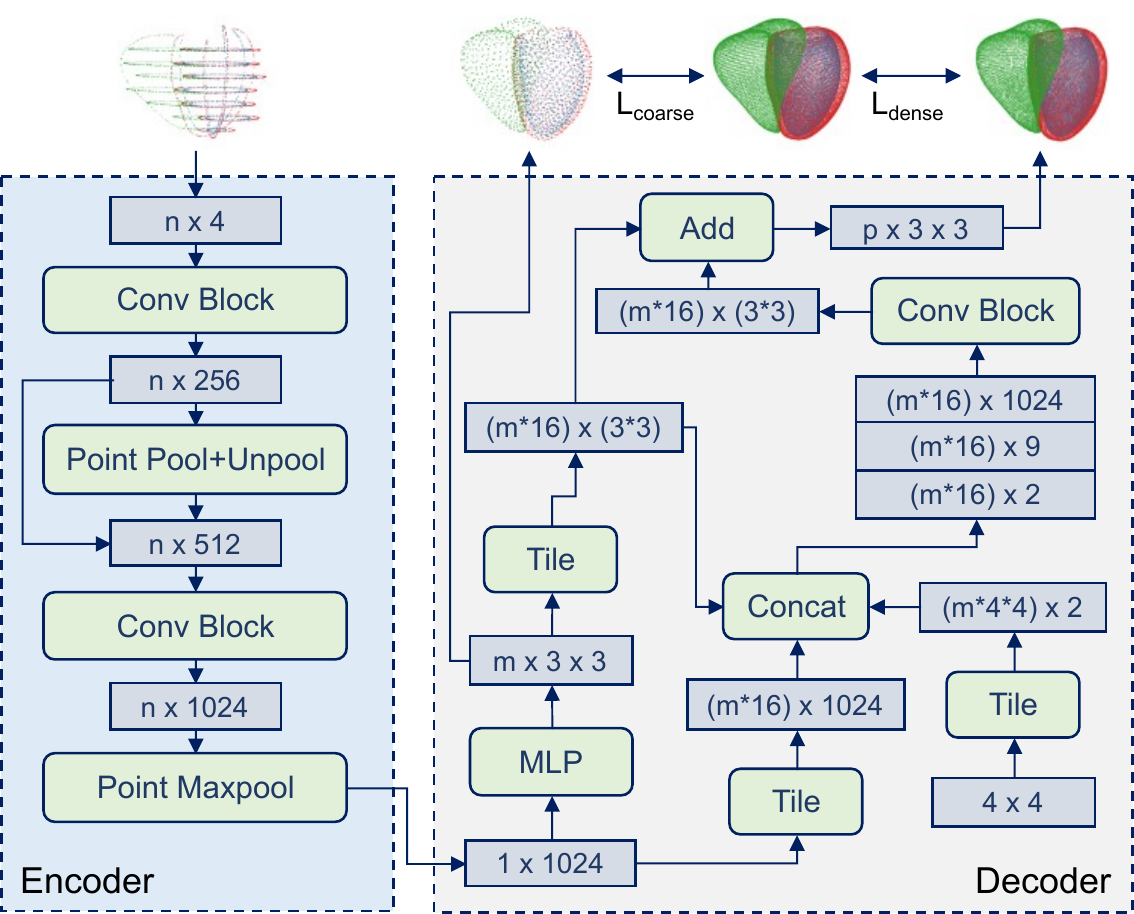}}
	\caption{Architecture of the proposed point cloud completion network. The input is a 3D point cloud, which represents the sparse and misaligned cine MRI acquisitions as a $n \times 4$ tensor where $n$ refers to the number of points and $4$ to the spatial $x,y,z$ coordinates with a class label for each point. The network is tasked to reconstruct both a coarse, low-density point cloud to capture the global surface structure and a dense, high resolution point cloud to accurately represent the cardiac anatomy on both a local and global level and serve as the final network prediction. The three anatomical substructures are encoded as separate sets of $x,y,z$ point coordinates in each of the output point clouds. Accordingly, the output dimensionality is $m \times 3 \times 3$ for the coarse point cloud and $p \times 3 \times 3$ for the dense point cloud where $m$ and $p$ refer to the respective number of points.}
	\label{fig:pcn_architecture}
\end{figure}

The PCCN architecture is based on recent advances in point cloud-based deep learning, in particular PointNet \citep{qi2017pointnet}, PointNet++ \citep{qi2017pointnet++}, FoldingNet \citep{yang2017foldingnet}, and Point Completion Network \citep{yuan2018pcn} (Fig.~\ref{fig:pcn_architecture}).

It consists of an encoder-decoder structure with a latent space vector of size 1024. The encoding part of the network is an adapted version of PointNet \citep{qi2017pointnet} to allow multi-class point cloud processing with different resolutions for input and output data. The network input is a sparse, misaligned point cloud of size $n \times 4$, where a scalar class variable to identify the cardiac substructure (LV cavity, LV myocardium, RV cavity) is concatenated to the $3$ spatial coordinate values $(x,y,z)$ of each of the $n$ points. Inspired by the design of PointNet++ \citep{qi2017pointnet++}, the input is fed through two combinations of PointNet-style \citep{qi2017pointnet} convolutional blocks and pooling operations as well as a skip connection to allow the network to access information at different scales and across per-point feature maps, before passing the output vector to the decoder.

The decoder architecture exhibits a similar two-step design as the decoder of the Point Completion Network \citep{yuan2018pcn}, but is also adapted to our high-density and multi-class setting. The first step is inspired by \citet{achlioptas2018learning} and inputs the latent space vector into a shared multilayer perceptron (MLP) followed by a reshaping operation to generate a coarse 3D point cloud with $m$ points separately for each of the three anatomical classes. The goal of this low-resolution point cloud is to capture the global shape of the biventricular anatomy by distributing the 3D points along the surfaces of the respective anatomical structures so that the highest-possible coverage is achieved. The second part of the decoder is based on FoldingNet \citep{yang2017foldingnet} where points are first initialized as grid-structured patches of size $4 \times 4$ with the tiling operation where each patch corresponds to one of the points in the coarse 3D point cloud and is then iteratively deformed to obtain the best-possible fit with the dense target surface of the ground truth point cloud. This leads to an effective increase in point cloud resolution on a local level to obtain the final dense output point cloud while maintaining the global information of the coarse point cloud output. The size of the final dense output point cloud is $p \times 3 \times 3$, where $p$ refers to the number of points, the first $3$ to the spatial coordinates, and last $3$ to the respective cardiac substructures. In this work, we set $n$, $m$, and $p$ to 36000, 750, and 12000, respectively.

\subsubsection{Loss function and training}
\label{sec:loss_function}

We base the loss function to train our PCCN on \citet{yuan2018pcn} and extend it to a multi-class setting by summing over the loss values of each class to obtain a combined total loss. The class-specific loss function consists of two loss terms defined at two different stages of the decoder path as
\begin{equation}
L_{total} = \displaystyle\sum_{i = 1}^{C} \big(L_{coarse, i} + \alpha * L_{dense, i}\big)
\label{eq_loss}
\end{equation}
where $C$ refers to the number of classes in the biventricular anatomy. Although we have only tested $C=3$ in this work, the proposed approach can be easily extended to any number of classes.
The first loss term $L_{coarse}$ compares the coarse 3D point cloud after the first decoder step with the dense ground truth point cloud and forces the sparse, intermediate point cloud to be a good representation of the global shape. The second loss term $L_{dense}$ acts on the final high-resolution point cloud prediction and enforces the desired smooth shape representation on both a global and local level. The weight $\alpha$ is used to control the importance of each of the two loss terms in the total loss. We choose a low $\alpha$ of 0.01 at the beginning of training to allow the network to first learn a good coarse representation of the global anatomy. As training progresses, $\alpha$ is gradually increased to focus on local anatomical details. We use the Chamfer distance between reconstructed and ground truth point clouds for both loss terms in (\ref{eq_loss}):
\begin{equation}
\begin{aligned}
CD(P_{1}, P_{2}) = \frac{1}{2} \bigg( \frac{1}{|P_{1}|}\displaystyle\sum_{x \in P_{1}} \min_{y \in P_{2}} \| x - y \|_{2} + \\
\frac{1}{|P_{2}|}\displaystyle\sum_{y \in P_{2}} \min_{x \in P_{1}} \| y - x \|_{2} \bigg)
\end{aligned}
\label{eq_acd}
\end{equation}
where $P_{1}$ refers to the predicted point cloud and $P_{2}$ to the ground truth point cloud.

We train the network for $2000$ epochs on a GeForce RTX 2070 Graphics Card using the Adam optimizer \citep{kingma2015adam} and a batch size of 8. The learning rate is initially set to 0.0001 and reduced every $30k$ steps with a decay rate of 0.7 to enable finer network updates as training progresses.

\subsubsection{Synthetic dataset generation}
\label{sec:dataset_generation}

Since we do not have access to a large number of ground truth 3D anatomies, we construct a synthetic dataset from the statistical shape model (SSM) described in Sec.~\ref{sec:statistical_shape_model} to train our PCCN. 

\begin{figure}[!t]
	\centerline{\includegraphics[width=0.485\textwidth]{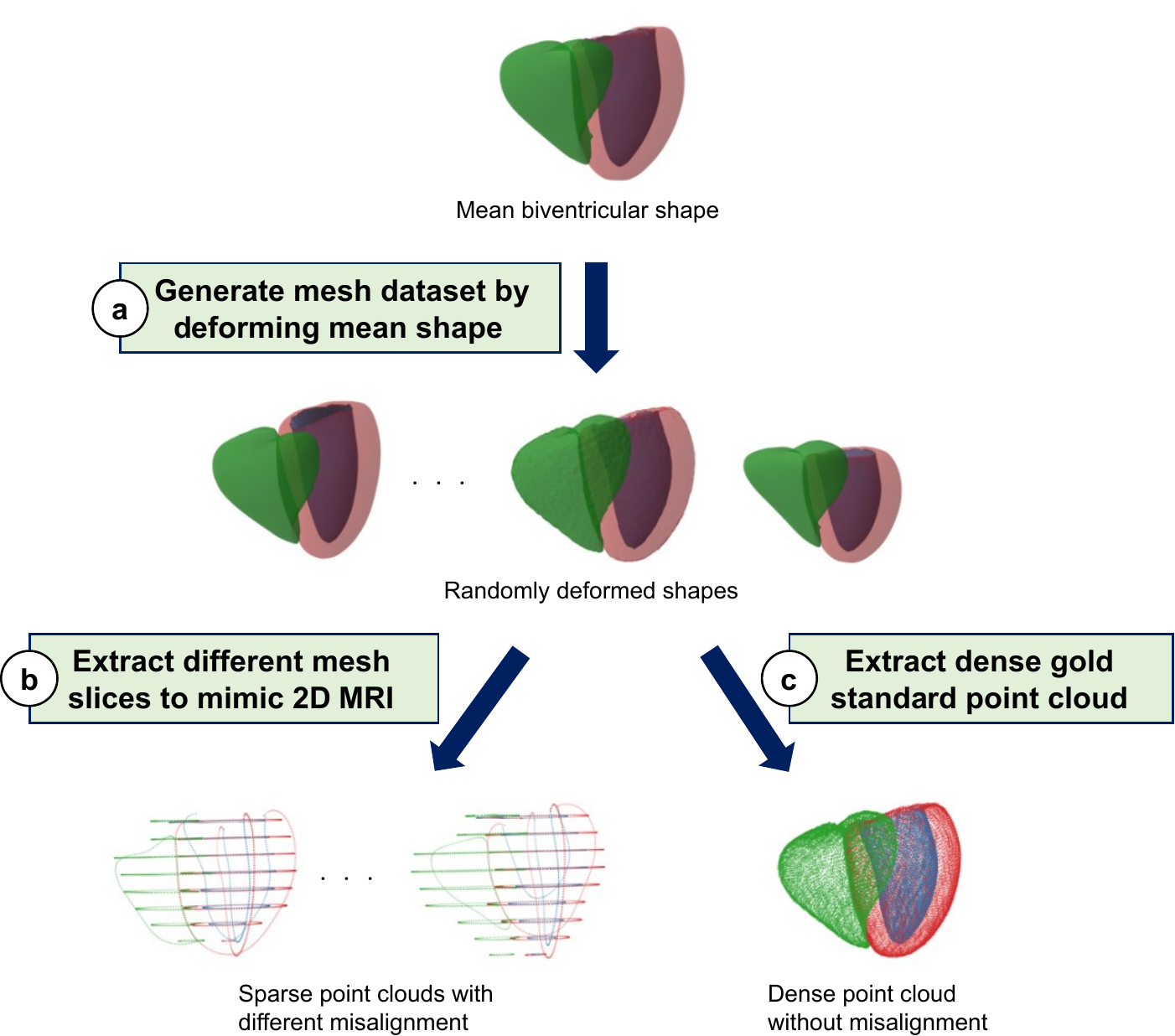}}
	\caption{Overview of the synthetic dataset generation from a 3D MRI-based statistical shape model.}
	\label{fig:dataset_generation}
\end{figure}

To this end, we design a three-step process to synthesize the sparse and misaligned input point clouds as well as the corresponding dense ground truth point clouds (Fig.~\ref{fig:dataset_generation}). First, we generate a virtual population of biventricular anatomy meshes by sampling from the SSM (Fig.~\ref{fig:dataset_generation}-a). Next, we determine the slice planes of the generated meshes that best mimic the clinically standard cine MRI acquisition \citep{taylor2005cardiovascular,walsh2007assessment,margeta2014recognizing}. We introduce small random translations to the chosen landmark points to recreate possible human errors during acquisition.
We then artificially introduce misalignment artifacts due to respiratory and patient motion to each SAX and LAX slice, allowing us train the PCCN under realistic conditions. Hereby, we assume that the misalignment can be fully described by rigid transformations, which we found to be a good approximation of real conditions. In order to introduce different random misalignments for each case, we sample the transformation parameters from a normal distribution with zero mean separately for the $x,y,z$ translations and the rotations around the $x,y,z$-axes respectively. To systematically analyze the performance of our method for different misalignment amounts, we introduce five subgroups with different average levels of randomly introduced misalignment, starting with no misalignment and then increasing the misalignment amount for each level (mild, medium, strong, severe). We choose five separate normal distributions with increasing standard deviation values for each of the subgroups to induce said differences in average misalignment between the different subgroups. Following previous pertinent literature \citep{mcleish2002study,shechter2004respiratory,chandler2008correction,villard2016correction,xu2019ventricle,tarroni2020large} (see Supplementary Material for more details), we select the standard deviation values in Table~\ref{table:misalignment_amounts} as reasonable approximations for typical misalignment amounts found in real acquisitions for each severity level.

We introduce random misalignment in this way to each slice of the whole SAX stack and both LAX slices before converting the slices into 3D point clouds, which now represent the sparse, misaligned cine MRI contours of a realistic acquisition (Fig.~\ref{fig:dataset_generation}-b). Finally, we extract the vertices of the corresponding deformed meshes generated from the SSM to obtain the dense ground truth point clouds for network training (Fig.~\ref{fig:dataset_generation}-c).

We run this 3-step synthetic dataset generation process (Fig.~\ref{fig:dataset_generation}) separately for each of the five levels of misalignment to create five different SSM-based datasets. For each of the four datasets with slice misalignment (mild, medium, strong, severe), we first generate 250 deformed meshes for both ED and ES phases and then apply 10 different sets of random misalignment transformations to each of the meshes, resulting in 5000 sparse, misaligned point clouds per misalignment level. In case of no misalignment, we sample 500 different shapes from the SSM for both ED and ES phases, and apply 5 different random transformations to mimic errors in slice plane selection to each of the 1000 point clouds. We note that no individual correspondence between generated ED and ES shapes is present in the dataset based on the available SSM data. Each of the five datasets is split into train, validation, and test datasets with sizes $80\%$, $5\%$, and $15\%$, respectively.

\begin{table}
	\caption{Misalignment amounts per severity level.}
	\label{table:misalignment_amounts}
	{\def\arraystretch{1.5}\tabcolsep=3pt
        \resizebox{0.49\textwidth}{!}{
		\begin{tabular}{|@{ }l@{ }|p{25pt}|p{25pt}|@{ }l@{ }|@{ }l@{ }|p{30pt}| } \hline
			{} &  \multicolumn{5}{c|}{Misalignment Level}\\ \cline{2-6}
			{}   &	None & Mild & Medium & Strong & Severe\\ \hline
			Translation (mm) & 0.0 & 1.5 & 2.5 & 3.5 & 5.0 \\ \hline
			Rotation ($^\circ$) & 0.0 & 0.5 & 1.5 & 2.5 & 3.5 \\ \hline
			\multicolumn{6}{p{240pt}}{Standard deviation values of normal distributions with zero mean for each level of misalignment.}
		\end{tabular}
        }}
\end{table}

\subsection{Surface mesh generation from dense point cloud}
\label{sec:meshing_step}

The last step of our surface reconstruction pipeline consists of transforming the multi-class biventricular point clouds into triangular meshes. To this end, we select the Ball Pivoting algorithm \citep{bernardini1999ball} and apply it separately for each of the three cardiac substructures of the reconstructed point clouds. This allows us to use different hyperparameter settings in the meshing algorithm for each class to account for their specific topological requirements.

\section{Experiments}
\label{sec:experiments}

In this section, we first evaluate our proposed point cloud completion network on the SSM dataset (Sec.~\ref{sec:experiments_ssm}) and compare its performance to a 3D U-Net benchmark (Sec.~\ref{sec:experiments_unet_comp}). We then validate the complete cardiac surface reconstruction pipeline on the UK Biobank dataset from both a geometric (Sec.~\ref{sec:experiments_ukbb}) and clinical perspective (Sec.~\ref{sec:experiments_clinical_metrics}) and analyze its robustness (Sec.~\ref{sec:experiments_robustness}).

\begin{figure}[!t]
        \centerline{\includegraphics[width=0.485\textwidth]{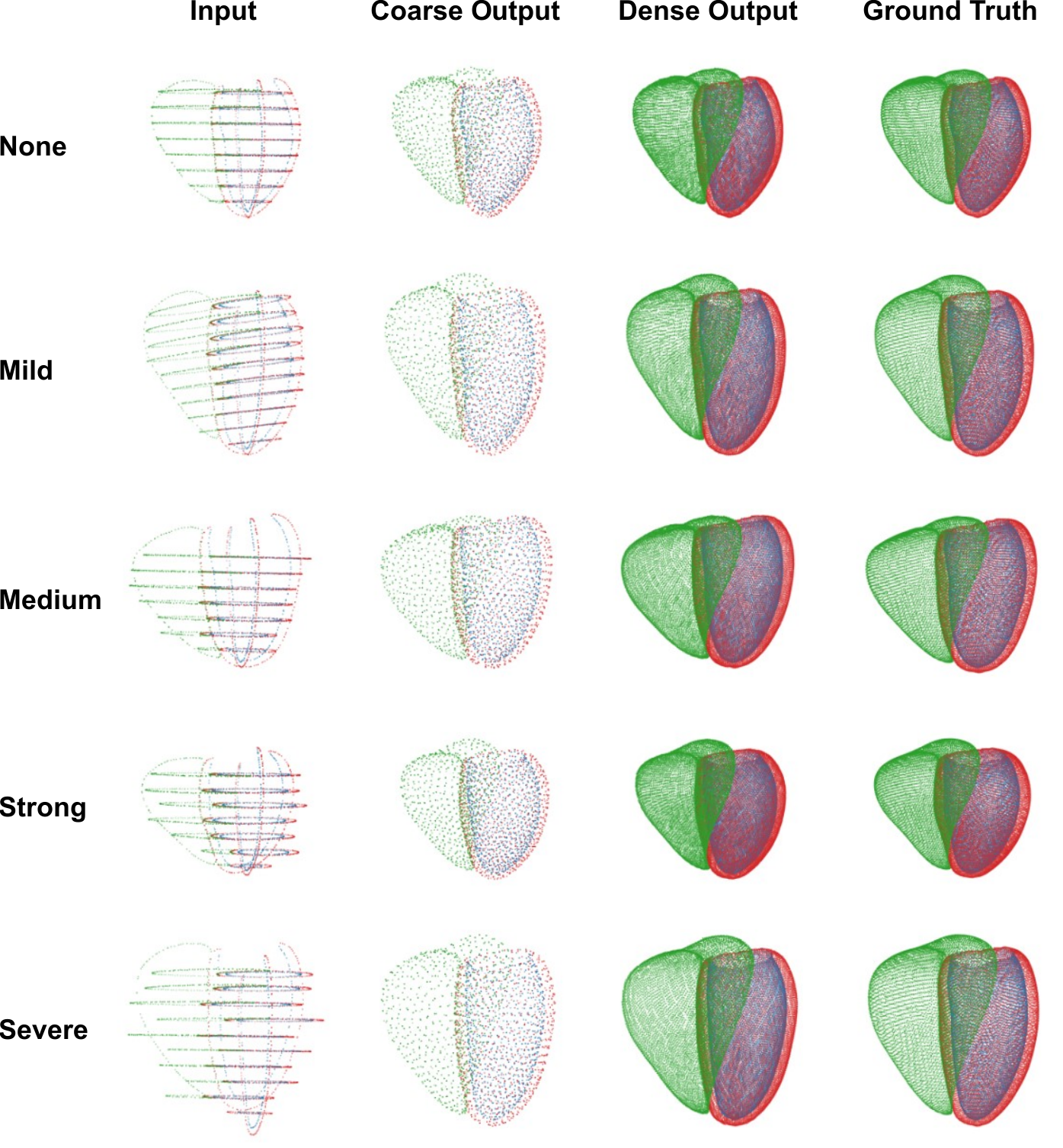}}

	\caption{Qualitative reconstruction results of an ED sample case for each of the five levels of misalignment from the SSM dataset.}
	\label{fig:ssm_reconstruction_results_ED}
\end{figure}

\subsection{Statistical shape model dataset}
\label{sec:experiments_ssm}

\begin{figure}[!t]
	\centerline{\includegraphics[width=0.485\textwidth]{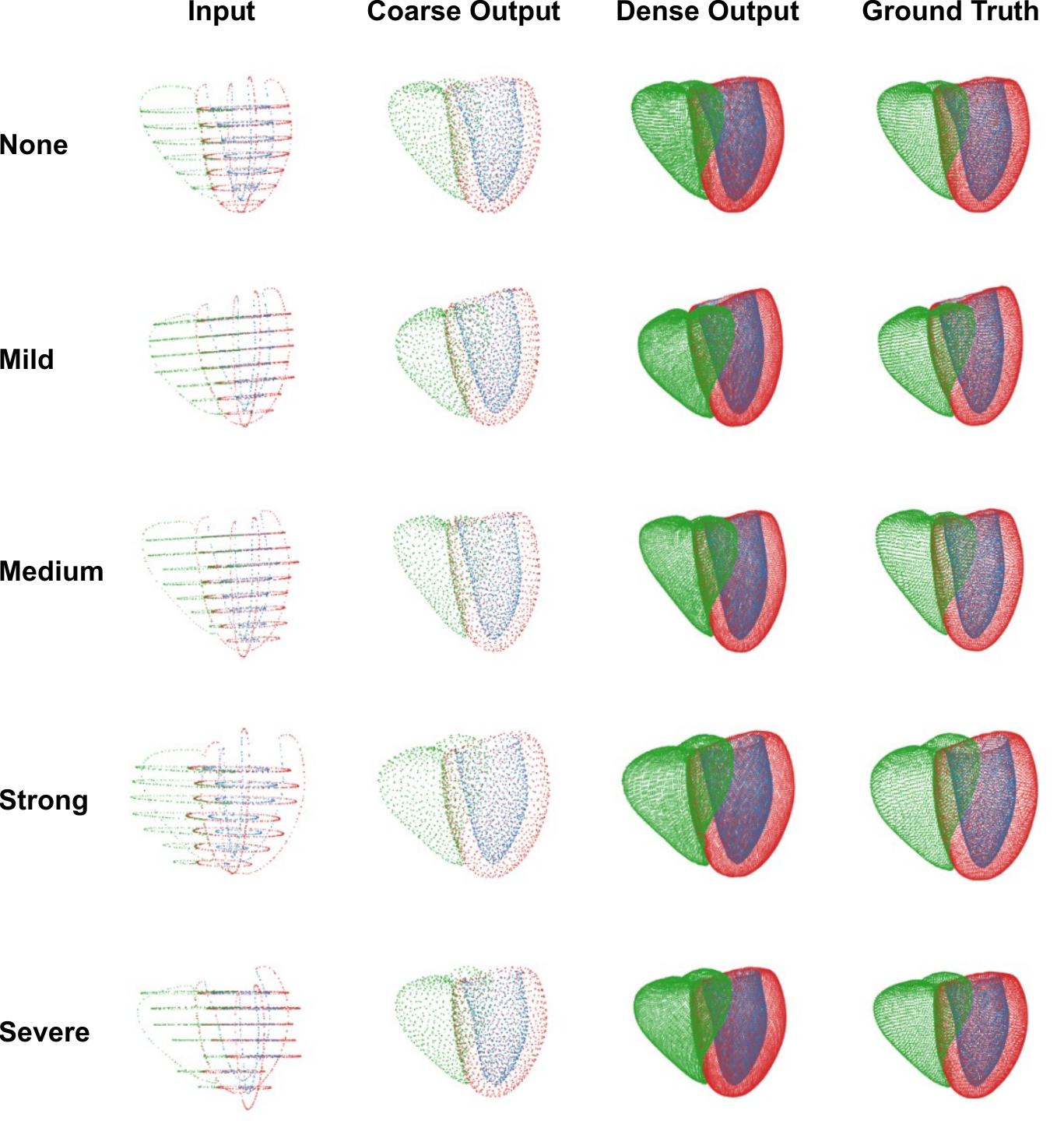}}
	\caption{Qualitative reconstruction results of an ES sample case for each of the five levels of misalignment from the SSM dataset.}
	\label{fig:ssm_reconstruction_results_ES}
\end{figure}

We choose the synthetic SSM dataset (Sec.~\ref{sec:dataset_generation}) for the first evaluation of our point cloud completion network, as it enables a direct comparison between the available ground truth anatomies and the reconstructed point clouds and meshes. By introducing slice misalignment at five different levels of severity, we can also analyze the effect of misalignment on the performance of the network. To this end, we train five different networks on each of the training datasets that correspond to one of the five levels of misalignment and validate our method on the respective unseen test datasets. For each misalignment level, both ED and ES point clouds are included in the respective datasets allowing an analysis of the network's reconstruction performance on multi-temporal data. Figures~\ref{fig:ssm_reconstruction_results_ED} and \ref{fig:ssm_reconstruction_results_ES} depict the sparse, misaligned input point cloud, both network outputs, and the pertinent ground truth point cloud per misalignment level for multiple different ED and ES sample cases, respectively.

We observe that the network is able to reconstruct the biventricular anatomy with high accuracy for a variety of different shapes and sizes on both a local and global level. Reconstruction quality decreases slightly as the amount of misalignment in the input point clouds increases. The basal areas of the cardiac anatomy show the most disagreement between prediction and ground truth, due to the high information sparseness in the input point clouds in this region. The different cardiac substructures and the two cardiac phases perform similarly well in the reconstruction task.

\begin{figure*}[!t]
\centerline{\includegraphics[width=0.9\textwidth]{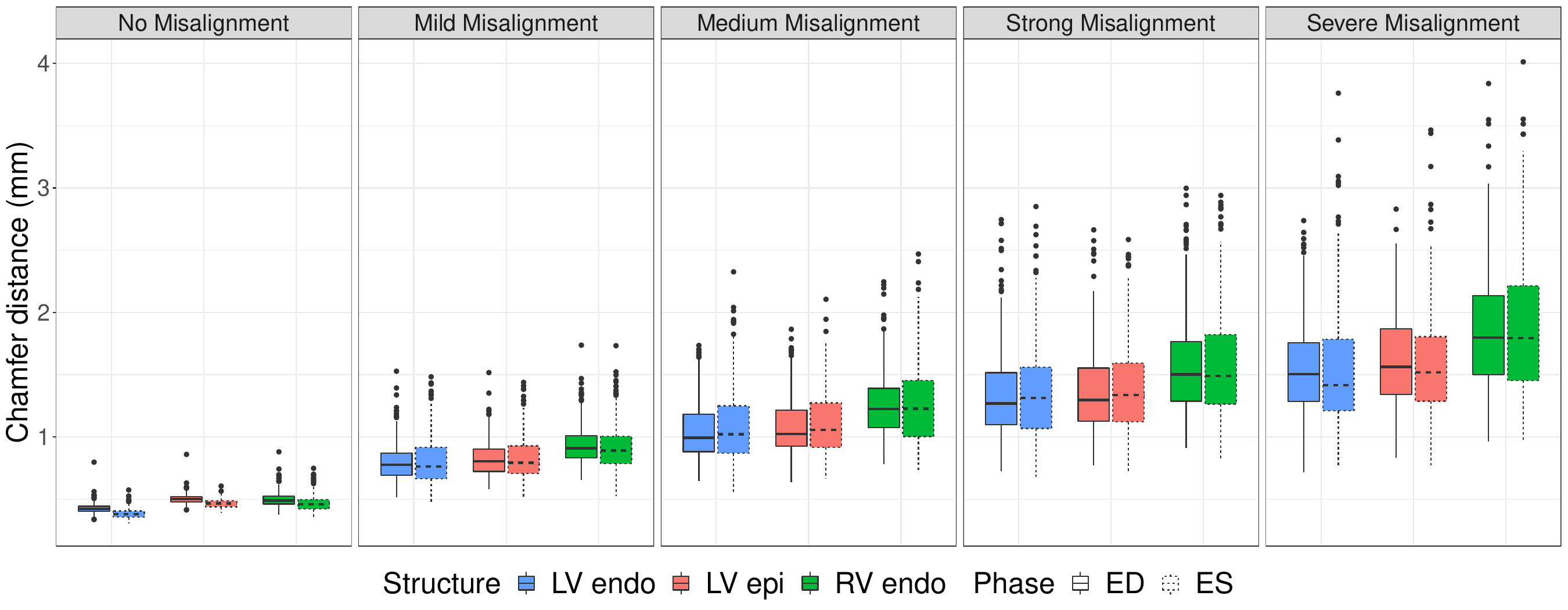}}
\caption{Boxplots presenting Chamfer distances between ground truth and reconstruction results of our method on five different SSM datasets with increasing levels of slice misalignment.}
\label{fig:misalignment_levels_recon_results_ssm}
\end{figure*}

In order to quantify the reconstruction ability of our point cloud completion network for different misalignment amounts, we calculate the Chamfer distances between the dense predicted point clouds and the corresponding ground truth point clouds in each of the five test datasets. We report the results in Fig.~\ref{fig:misalignment_levels_recon_results_ssm}, split by cardiac substructure and phase for each of the five levels of slice misalignment.


We find median Chamfer distances considerably below or close to the underlying image resolution ($1.8 \times 1.8$~mm) with low quartile deviation values for all levels of misalignment, cardiac phases, and substructures. Both median and quartile deviation values increase with rising levels of misalignment. Chamfer distances are generally higher for the right ventricular anatomies than for the left ventricular ones, while only marginal differences generally exist between the ED and ES phases.

\subsection{Comparative analysis}
\label{sec:experiments_unet_comp}

We compare our PCCN with a state-of-the-art 3D U-Net architecture \citep{cciccek20163d}, which has previously been applied to biventricular surface reconstruction \citep{xu2019ventricle}. For this task, we select the SSM dataset with medium misalignment as it represents the mean slice misalignment expected in a typical cine MRI acquisition \citep{mcleish2002study,shechter2004respiratory,chandler2008correction,xu2019ventricle,tarroni2020large,villard2016correction}.

\begin{figure}[!t]
		\centerline{\includegraphics[width=0.485\textwidth]{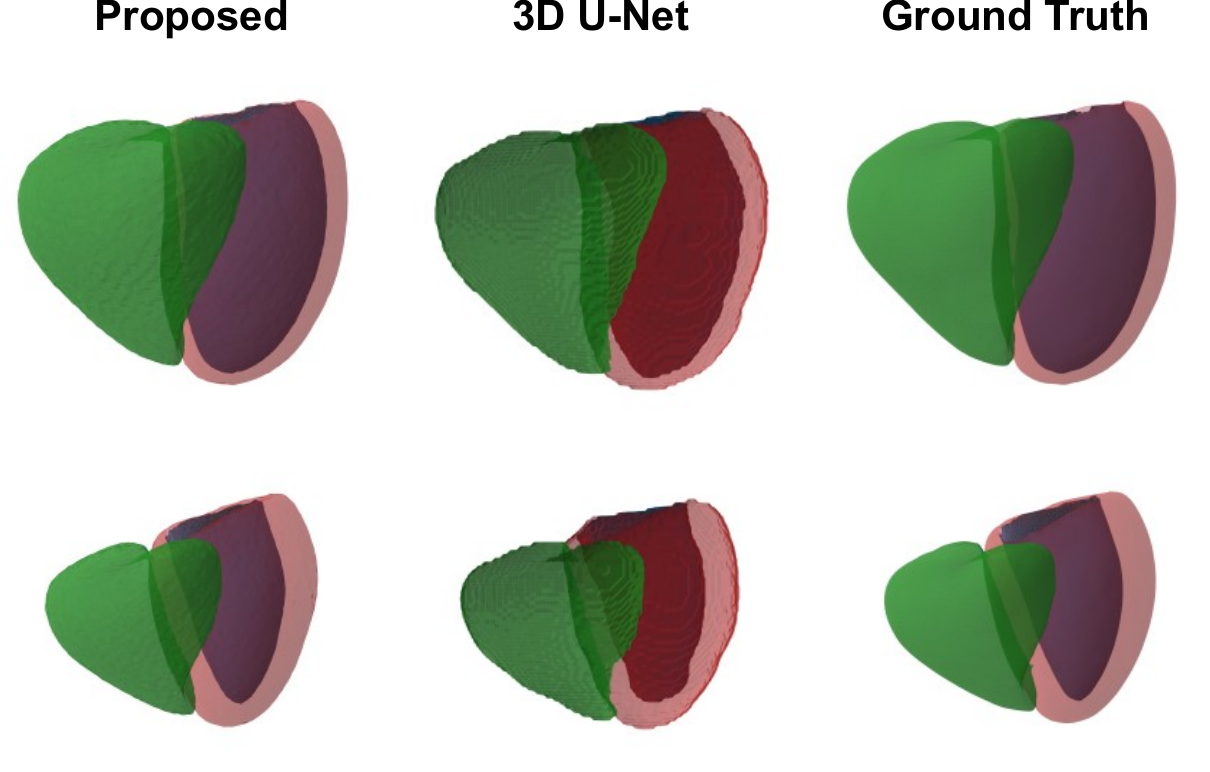}}

	\caption{Qualitative reconstruction results on two sample cases of the SSM dataset for the Point Cloud Completion Network and a 3D U-Net.}
	\label{fig:method_comparison_results_ssm}
\end{figure}

Since U-Nets operate on grid-based structures, we first convert the sparse, misaligned input point clouds and the ground truth point clouds of our dataset to voxel grid representations of the biventricular anatomy. We set the voxel size to $1.5 \times 1.5 \times 1.5$~mm, chosen as a trade-off between closeness to the 3D MRI resolution underpinning the SSM dataset ($1.25 \times 1.25 \times 2$~mm) and ensuring that the complete biventricular anatomy fits into the fixed size $128 \times 128 \times 128$ voxel grid for all cases. It is also smaller than the pixel size of the underlying 2D image acquisition ($1.8 \times 1.8$~mm) which acts as a lower accuracy limit of the point cloud representation. Furthermore, the voxel resolution values are slightly higher than the $2 \times 2 \times 2$~mm used in the work by \citet{xu2019ventricle}, enabling a more accurate reconstruction.

With both point cloud and voxel grid representations available for each case, we train both our PCCN and a 3D U-Net for biventricular surface reconstruction on the same dataset. To allow a comparison of results, which is as afair as possible, we convert both the point clouds and voxel grids to multi-class triangular meshes as a neutral data type by using the Ball Pivoting \citep{bernardini1999ball} and Marching Cubes algorithms \citep{lorensen1987marching}, respectively. The resulting meshes predicted by the PCCN and the 3D U-Net as well as the corresponding ground truth meshes are shown for two sample cases in Fig.~\ref{fig:method_comparison_results_ssm}.

We observe that both the PCCN and the U-Net are able to accurately reconstruct different cardiac shapes for all cardiac substructures and phases. On a global level, we notice only minor differences between the results, which are mostly caused by the lower smoothness of the U-Net outputs as a result of deriving them from gridded data. Visible differences are larger on a local level where the U-Net reconstructions exhibit erroneous outward bulging in some surface regions that do not align with the ground truth and are correctly smoothed out in the respective PCCN predictions. These differences most commonly occur in the LV cavity substructure and are slightly more pronounced in ES than in ED.

In order to quantify the differences between the PCCN and the 3D U-Net, we calculate the Hausdorff distances, the mean surface distances (MSD), and the Chamfer distances between predicted and ground truth meshes of the unseen SSM test dataset for both methods and report the results in Table~\ref{tab:quant_method_comp_ssm}. In addition, we also provide information about the number of network parameters and data representations used in the respective approaches.

We find that the PCCN outperforms the 3D U-Net by 32\% and 24\% in terms of average Hausdorff distance and mean surface distance, respectively. Standard deviations of both distance metrics are also lower for the PCCN than for the U-Net reconstructions, while only minor differences exist between the ED and ES results. Due to its usage of memory-efficient point clouds, the PCCN achieves this outperformance despite using 13 times less storage space for each anatomy.

\begin{table}[!ht]
	\caption{Comparison of cardiac mesh reconstruction methods using the SSM dataset with medium misalignment. The distance scores are averaged across the three cardiac substructures.}
	\label{tab:quant_method_comp_ssm}
	{\def\arraystretch{1.5}\tabcolsep=3pt
            \resizebox{0.49\textwidth}{!}{
	    \begin{threeparttable}
		\begin{tabular}{|p{100pt}|p{60pt}|p{60pt}|}
			\hline
			{}   &	Proposed & 3D U-Net$^{1}$ \\
			\hline
			Data type & point cloud & voxel grid \\
			\hline
			Input data size & $\sim$144 $\times$ 10$^3$ & $\sim$2 $\times$ 10$^6$ \\
			\hline
			Output data size & $\sim$108 $\times$ 10$^3$ & $\sim$2 $\times$ 10$^6$ \\
   			\hline
			Number of parameters & $\sim$10.6 $\times$ 10$^6$ & $\sim$12.0 $\times$ 10$^6$ \\
			\hline
			ED Hausdorff (mm)$^2$ & 3.50 $\pm$ 0.84 & 5.31 $\pm$ 1.63 \\
			\hline
			ES Hausdorff (mm)$^2$ & 3.49 $\pm$ 0.89 & 4.83 $\pm$ 1.09 \\
			\hline
			ED MSD (mm)$^2$ & 0.93 $\pm$ 0.28 & 1.30 $\pm$ 0.44 \\
			\hline
			ES MSD (mm)$^2$ & 0.98 $\pm$ 0.32 & 1.21 $\pm$ 0.36 \\
                \hline
			ED Chamfer (mm)$^2$ & 1.13 $\pm$ 0.26 & 1.49 $\pm$ 0.46 \\
			\hline
			ES Chamfer (mm)$^2$ & 1.15 $\pm$ 0.29 & 1.30 $\pm$ 0.35\\
			\hline
		\end{tabular}
	    \begin{tablenotes}
	        \item[1] \citet{cciccek20163d}.
            \item[2] Values represent \textit{mean $\pm$ SD}.
        \end{tablenotes}
        \end{threeparttable}
        }}
\end{table}

\begin{figure*}[!t]
\centerline{\includegraphics[width=\textwidth]{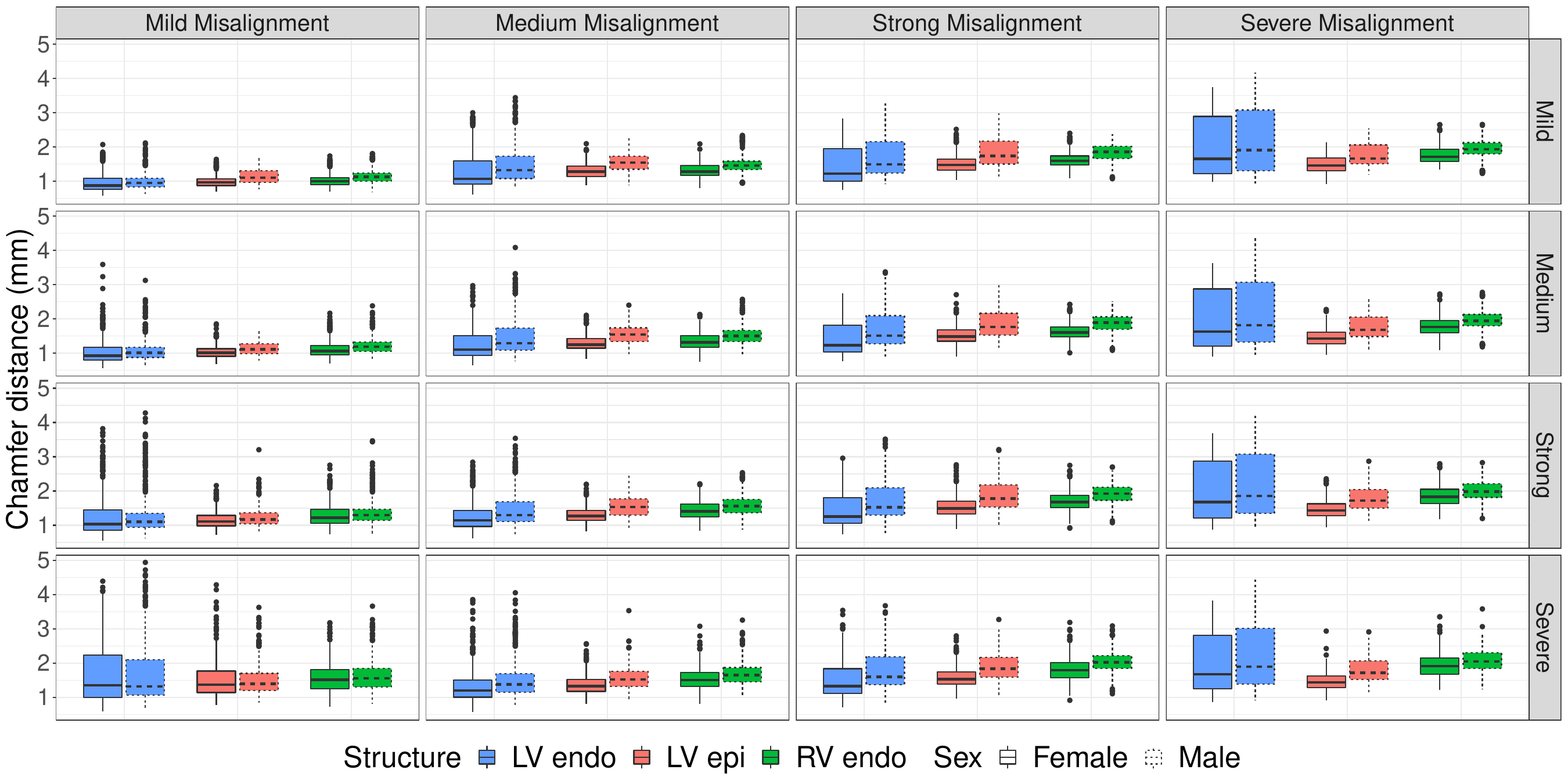}}
\caption{Boxplots showing the reconstruction performance of networks trained on SSM datasets with different misalignment levels (in columns) and applied to UKB data with different misalignment levels (in rows).}
\label{fig:quant_recon_results_ukbb}
\end{figure*}

\subsection{UK Biobank}
\label{sec:experiments_ukbb}

After evaluating our PCCN on the synthetic SSM datasets, we assess the ability of the complete cardiac surface reconstruction pipeline to transform raw cine MR images into triangular mesh representations of the biventricular anatomy on the real-world dataset of the UK Biobank study \citep{petersen2013imaging,petersen2015uk}. Since the first two steps of our 4-step reconstruction pipeline were both developed and validated on the MRI acquistions of the UK Biobank \citep{bai2018automated,banerjee2021ptrsa}, we can directly use them for this task without any further adjustments. As the third step of our pipeline, we want to directly apply a PCCN pre-trained on one of the SSM datasets to the sparse, misaligned UK Biobank point clouds obtained in the second pipeline step in a cross-domain transfer setup. To this end, we first assess which of the five networks, trained on different amounts of misalignment, is the best fit for the UK Biobank dataset. We refer to these networks as no, mild, medium, strong, or severe misalignment networks for the remainder of this paper.

Since this is not a straightforward task due to the lack of available 3D ground truth for the UK Biobank data, we create a set of approximate ground truth meshes to act as a benchmark for our analysis. We first select 10 cases with the least amount of misalignment in the UK Biobank dataset. Hereby, we determine the misalignment amount of each case by calculating the average shortest distance of each point in each slice to the remaining slices in the given point cloud. The corresponding 3D point clouds are then reconstructed for both ED and ES using the PCCN trained on the SSM dataset with no misalignment. We consider these 3D reconstructions as our pseudo-gold standard for this experiment. However, we note that some reconstruction error is still expected to be present, as the selected cases are not completely without misalignment, come from a different, unseen domain compared to the PCCN's training SSM dataset, and might contain segmentation errors.

Given this set of pseudo ground truth anatomies, we can apply each of the four pre-trained PCCN candidate models to the sparse input point clouds and compare the predicted 3D reconstructions with the corresponding pseudo ground truths. However, this would only assess the performance on UK Biobank cases with very little misalignment, which are not representative of the whole dataset. Hence, we first artificially introduce random slice misalignments to each of the sparse input point clouds to mimic real-world misalignment conditions, while still maintaining our pseudo gold standard point clouds required for the comparative evaluation. Similar to our experiments on the SSM dataset, we include both ED and ES point clouds of each case in the dataset and introduce the misalignment at four different levels of severity (mild, medium, strong, severe). For each level, 10 random amounts of misalignment are applied to each of the 10 pseudo ground truth cases, resulting in 100 misaligned and sparse point clouds per misalignment level. We use the Chamfer distances between the predicted and pseudo gold standard point clouds as our evaluation metric in all cases and report the UK Biobank reconstruction results separated by cardiac substructure and sex in Fig.~\ref{fig:quant_recon_results_ukbb}.

We observe that the mild misaligment PCCN generally achieves the best overall results across all misalignment levels, cardiac substructures, and sex. Its distance scores are the lowest for mild, medium, and strong UKB misalignments, as well as for severe UKB misalignment in the RV endocardium. The medium misalignment PCCN performs best on severely misaligned left ventricular UKB data and second-best overall. We also see a general decrease in performance of all four analyzed networks with increasing misalignment in the UKB data.

\begin{figure}[!t]
		\centerline{\includegraphics[width=0.485\textwidth]{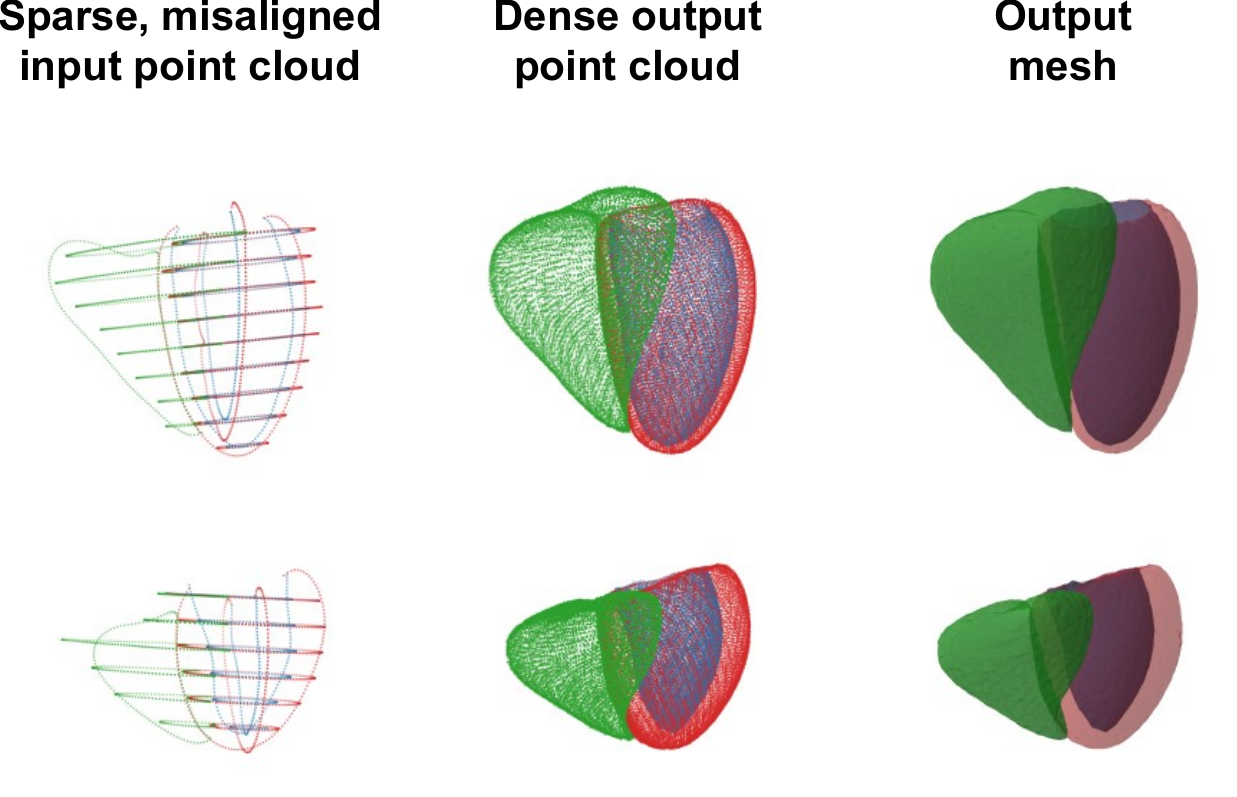}}

	\caption{Qualitative reconstruction results for two sample cases of the UK Biobank dataset.}
	\label{fig:UKB_sample_results}
\end{figure}

Based on these quantitative evaluation results on the UK Biobank data, we select the mild misalignment PCCN for the third step of our reconstruction pipeline.
With all components of the full reconstruction pipeline available, we apply it to the randomly selected 1000 subjects of the UK Biobank dataset. We visualize the sparse, misaligned input point clouds, the corresponding dense output point clouds, and the output meshes for two sample UK Biobank cases in Fig.~\ref{fig:UKB_sample_results}.

We find realistic and plausible 3D reconstructions that align very well with the 2D anatomical information in the sparse input point clouds for all cardiac substructures and phases. Furthermore, the meshing step is able to successfully preserve the cardiac surface anatomy of the reconstructed point clouds in the final output meshes and create topologically accurate two-manifold meshes for ~97\% of all cases. Only small differences in reconstruction performance between the ED and ES phases are observed.

\subsection{Clinical metrics}
\label{sec:experiments_clinical_metrics}

Next, we evaluate the ability of our cardiac surface reconstruction method to generate clinically plausible meshes on a population level. To this end, we select two population-wide studies of healthy cardiac anatomy and function and compare their results with ours in terms of multiple clinically established cardiac image-based biomarkers. Table~\ref{tab:reference_studies_overview} provides an overview of the two benchmark studies along with our proposed method.

\begin{table}[!t]
    \caption{\label{tab:reference_studies_overview}Dataset comparison of the proposed and benchmark studies.}
    \def\arraystretch{1.5}\tabcolsep=3pt
        \centering
        \resizebox{0.49\textwidth}{!}{
        \begin{tabular}{|p{45pt}|p{55pt}|p{70pt}|p{55pt}|}
            \hline
            {} & \citet{petersen2017reference} & \citet{bai2015bi} & Proposed \\
            \hline
            Dataset source    & UK Biobank  & UK Digital Heart Project  & UK Biobank   \\
            \hline
            Imaging type        & 2D MRI  & 3D MRI  & 2D MRI   \\
            \hline
            Resolution (mm)  & $1.8\times1.8\times8.0$  &  $1.25\times1.25\times2.0$  & $1.8\times1.8\times8.0$  \\
            \hline
            Slice gap (mm)          & 2.0  & None & 2.0   \\
            \hline
            Contouring procedure   & Manual  & Semi-automatic (subset manual)  & Fully automatic   \\
            \hline
            Biomarker calculation   & 2D-based  & 3D-based  & 3D-based   \\
            \hline
            Number of cases      &   800      & 1093  & 970  \\
            \hline
            Average age (yr)         &   59      &  40  & 62  \\
            \hline
    \end{tabular}
    }
\end{table}

\begin{table*}[!t]
    \caption{\label{tab:clinical_metrics_comp_per_sex}Clinical metrics of female and male cases reported by different studies.}
    \def\arraystretch{1.5}\tabcolsep=3pt
        \centering
        \resizebox{\textwidth}{!}{
        \begin{tabular}{|p{68pt}|p{50pt}|p{48pt}|p{45pt}|p{40pt}|p{50pt}|p{48pt}|p{45pt}|p{40pt}|}
            \hline
            {}      & \multicolumn{4}{c|}{Female}  & \multicolumn{4}{c|}{Male}  \\
			\cline{2-9}
            {}              & \citet{petersen2017reference}     & \citet{bai2015bi}     & Simpson's rule    & Proposed  & \citet{petersen2017reference}     & \citet{bai2015bi}     & Simpson's rule    & Proposed\\
    		\hline
			Number of cases  & 432                  & 600            & 500               & 483                  & 368               & 493           & 500           & 487\\
			\hline
			LVEDV (ml)       & 124 $\pm$ 21       & 138 $\pm$ 24    & 124 $\pm$ 22      & 128 $\pm$ 23      & 166 $\pm$ 32       & 178 $\pm$ 36              & 155 $\pm$ 30      & 156 $\pm$ 32 \\
			\hline
			LVESV (ml)       & 49 $\pm$ 11        & 48 $\pm$ 10     & 48 $\pm$ 12       & 50 $\pm$ 12       & 69 $\pm$ 16        & 65 $\pm$ 15               & 67 $\pm$ 19      & 67 $\pm$ 16\\
			\hline
			LVSV (ml)       & 75 $\pm$ 14        & -                & 75 $\pm$ 14       & 76 $\pm$ 16       & 96 $\pm$ 20        & -                        & 89 $\pm$ 18       & 88 $\pm$ 24\\
			\hline
			LV mass (g)     & 70 $\pm$ 13        & 96 $\pm$ 16      & 68 $\pm$ 12       & 82 $\pm$ 16       & 103 $\pm$ 21        & 128 $\pm$ 24             & 96 $\pm$ 17      & 120 $\pm$ 26\\
			\hline
			LVEF (\%)       & 61 $\pm$ 5        & 65 $\pm$ 4        & 61 $\pm$ 5        & 61 $\pm$ 6        & 58 $\pm$ 5        & 64 $\pm$ 4                 & 57 $\pm$ 6       & 57 $\pm$ 8\\
			\hline
			RVEDV (ml)      & 130 $\pm$ 24       & -                & 127 $\pm$ 24      & 148 $\pm$ 24      & 182 $\pm$ 36       & -                        & 167 $\pm$ 32      & 192 $\pm$ 32\\
			\hline
    		RVESV (ml)      & 55 $\pm$ 15       & -                 & 52 $\pm$ 13       & 64 $\pm$ 13       & 85 $\pm$ 22      & -                         & 76 $\pm$ 19        & 92 $\pm$ 18\\
			\hline
			RVSV (ml)       & 75 $\pm$ 14	    & -                 & 75 $\pm$ 15       & 84 $\pm$ 16       & 97 $\pm$ 20	    & -                         & 91 $\pm$ 19      & 99 $\pm$ 22\\
			\hline
			RVEF (\%)       & 58 $\pm$ 6       & -                  & 59 $\pm$ 6        & 58 $\pm$ 5        & 54 $\pm$ 6       & -                          & 55 $\pm$ 6        & 52 $\pm$ 7\\
			\hline
            \multicolumn{9}{@{}l}{Values represent \emph{mean $\pm$ standard deviation}.}
    \end{tabular}
    }
\end{table*}

We select the LV and RV volumes at both ED and ES phases as well as the LV myocardial mass as image-based biomarkers for the assessment of cardiac anatomy, while stroke volume (SV) and ejection fraction (EF) are used to quantify cardiac function for both the LV and RV. We calculate these metrics for all cases of our UK Biobank dataset using both the modified Simpson's rule on the 2D slice segmentations and the direct calculation from our reconstructed 3D meshes. The results are shown in Table~\ref{tab:clinical_metrics_comp_per_sex}, along with the corresponding values reported in the benchmark studies of \citet{petersen2017reference} and \citet{bai2015bi}. We split the scores by sex to analyze whether subpopulation-specific differences are accurately reflected in our method's reconstructions, providing additional validation of our proposed pipeline. We note that, while the analysis of \citet{petersen2017reference} is also based on the UK Biobank study, we use a different subset of cases in this work.

We observe that our 3D reconstruction pipeline achieves plausible scores for all analyzed metrics and is able to accurately capture sex-related differences. This is shown by the higher left and right ventricular volumes reported for male cases compared to the female ones, which is also present in all three benchmark studies. Comparing our 3D mesh-based approach with the two 2D slice-based calculation methods (Simpson's rule and \citet{petersen2017reference}), we find similar values for left ventricular volume (LV end-diastolic volume - LVEDV, LV end-systolic volume - LVESV) and function (LV stroke volume - LVSV, LV ejection fraction - LVEF) metrics, but larger values for LV mass and right ventricular volumetric metrics (RV end-diastolic volume - RVEDV, RV end-systolic volume - RVESV, RV stroke volume - RVSV). At the same time, our pipeline's scores are lower than the other 3D mesh-based approach by \citet{bai2015bi} in three out of the four available metrics and comparable for LVESV. The comparative analysis shows similar trends for both sexes with slightly larger differences for male cases.

In order to further analyze the ability of our method to take into account subpopulation-specific differences in its reconstruction task, we also calculate the same clinical metrics for three different age groups. The results of our 3D mesh-based calculations, the 2D slice-based approach using modified Simpson's rule, and the corresponding values reported by \citet{petersen2017reference} are reported for each of the three age groups in Table~\ref{tab:clinical_metrics_female_age_groups}. We only show the scores for female cases since the observed trends are similar for both sexes. The corresponding table for male cases can be found in the supplementary material.

\begin{table*}[!t]
    \caption{\label{tab:clinical_metrics_female_age_groups}Clinical metrics of female cases split by age group as reported by different studies.}
    \def\arraystretch{1.5}\tabcolsep=3pt
        \centering
        \resizebox{\textwidth}{!}{
        \begin{tabular}{|p{55pt}|p{50pt}|p{42pt}|p{40pt}|p{50pt}|p{42pt}|p{40pt}|p{50pt}|p{42pt}|p{40pt}|}
            \hline
            {}      & \multicolumn{9}{c|}{Age groups (years)}  \\
            \cline{2-10}
            {}      & \multicolumn{3}{c|}{45-54}  & \multicolumn{3}{c|}{55-64} & \multicolumn{3}{c|}{65-74}  \\
			\cline{2-10}
            {}      & \citet{petersen2017reference} & Simpson's rule    & Proposed  & \citet{petersen2017reference}  & Simpson's rule    & Proposed & \citet{petersen2017reference}  & Simpson's rule    & Proposed\\
			\hline
			LVEDV (ml)      & 131   & 132 $\pm$ 23      & 137 $\pm$ 24      & 121   & 127 $\pm$ 21      & 129 $\pm$ 22      & 122   & 119 $\pm$ 22      & 121 $\pm$ 21  \\
			\hline
			LVESV (ml)      & 52    & 52 $\pm$ 13       & 54 $\pm$ 13       & 47    & 49 $\pm$ 11      & 50 $\pm$ 11      & 48   & 46 $\pm$ 12      & 48 $\pm$ 11  \\
			\hline
			LVSV (ml)       & 79    & 79 $\pm$ 14      & 82 $\pm$ 17       & 74    & 77 $\pm$ 14      & 78 $\pm$ 17      & 74   & 72 $\pm$ 14      & 72 $\pm$ 15  \\
			\hline
			LV mass (g)     & 71   & 70 $\pm$ 11      & 80 $\pm$ 15       & 69   & 70 $\pm$ 12      & 83 $\pm$ 15      & 69   & 68 $\pm$ 11      & 81 $\pm$ 14  \\
			\hline
			LVEF (\%)       & 60    & 60 $\pm$ 5        & 60 $\pm$ 6        & 61    & 61 $\pm$ 5      & 61 $\pm$ 6      & 61   & 61 $\pm$ 6      & 60 $\pm$ 7  \\
			\hline
			RVEDV (ml)      & 138   & 134 $\pm$ 26      & 157 $\pm$ 29      & 125   & 129 $\pm$ 22      & 150 $\pm$ 23      & 128   & 121 $\pm$ 24      & 144 $\pm$ 22  \\
			\hline
    		RVESV (ml)      & 61    & 56 $\pm$ 14      & 67 $\pm$ 15       & 52    & 52 $\pm$ 12      & 64 $\pm$ 11      & 54   & 49 $\pm$ 13      & 62 $\pm$ 13  \\
			\hline
			RVSV (ml)       & 78    & 78 $\pm$ 15      & 89 $\pm$ 19       & 73    & 77 $\pm$ 14      & 86 $\pm$ 17      & 74   & 72 $\pm$ 15      & 81 $\pm$ 15  \\
			\hline
			RVEF (\%)       & 56    & 58 $\pm$ 5      & 57 $\pm$ 5        & 59    & 60 $\pm$ 5      & 57 $\pm$ 5      & 58   & 59 $\pm$ 7      & 57 $\pm$ 6  \\
			\hline
            \multicolumn{10}{@{}l}{Mean values are reported for \citet{petersen2017reference} and \emph{mean $\pm$ standard deviation} for Simpson's rule and proposed pipeline.}
    \end{tabular}
    }
\end{table*}

Similar to the sex-specific results, we find generally plausible scores for our 3D reconstructions and comparable trends between our 3D and the two 2D-based calculations with LV mass and RV metrics showing higher and the remaining metrics similar values. Our method is able to successfully capture clinically established age-related changes for all metrics. Examples include the decline in left and right ventricular volume at both ED and ES with increasing age and the consistent EF values across all age groups. In the former case, both our 3D and 2D-based calculations show decreases for both older age groups, while \citet{petersen2017reference} report small increases for the oldest age group compared to the medium one.

\subsection{Robustness analysis}
\label{sec:experiments_robustness}

To further validate the accuracy of our proposed reconstruction method on the UK Biobank dataset, we investigate its robustness to various common outlier conditions. In this regard, the image segmentation step of our pipeline is of considerable importance as it affects all downstream tasks, including the 3D surface reconstruction step with the PCCN. While the segmentation performance of modern deep learning approaches has generally been shown to be on par with human experts on a population level for healthy cases \citep{bai2018automated}, individual cases or slices often still result in erroneous outputs. These include the breakage of the LV myocardium in the apical region of the heart, the erroneous inclusion of papillary muscles in the myocardial region, anatomically incorrect segmentation of the basal plane slices, or the complete failure of the segmentation algorithm due to imaging artifacts which in turn results in missing slices in the 3D reconstruction task.

\begin{figure}[!ht]
		\centerline{\includegraphics[width=0.485\textwidth]{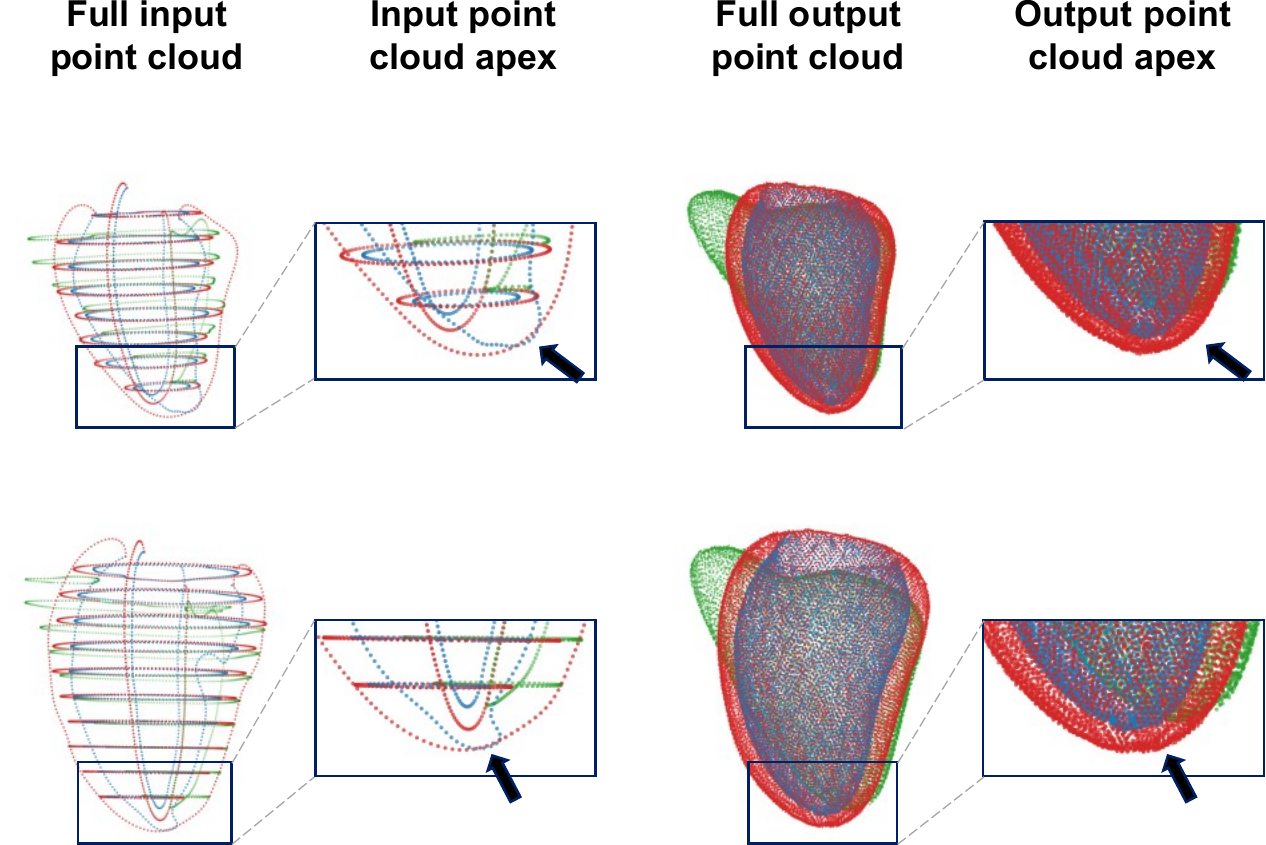}}

	\caption{Two sample cases with myocardial breakage in the apical region of the 2ch LAX segmentation mask.}
	\label{fig:samples_myocardial_breakage}
\end{figure}

In order to investigate the effects of such errors on the 3D surface reconstruction ability of the PCCN, we first select various UK Biobank cases that suffer from either myocardial breakage or erroneous segmentation of papillary muscles in the predicted segmentation masks. We then compare the affected regions in the sparse, misaligned input point clouds and the dense output point clouds reconstructed by the PCCN. The results are depicted for two sample cases of the UKB dataset in Fig.~\ref{fig:samples_myocardial_breakage}. 

We see that in both cases the PCCN is able to correct the myocardial breakage at the apex and reconstruct a smooth, continuous myocardium at the affected region. The bottom case in Fig.~\ref{fig:samples_myocardial_breakage} also depicts an erroneous segmentation of the papillary muscles, which are included in the myocardial region. This results in an inward bulging myocardium in the left mid-cavity region of the sparse input point cloud. However, similar to the myocardial breakage, the PCCN has successfully removed it from the dense output point cloud. We find this corrective ability of the PCCN present in all UK Biobank cases where either myocardial breakage or wrong papillary muscle segmentation occurs. In addition, the myocardial thickness in both reconstructed point clouds is smaller than suggested by the 2ch LAX view alone, but larger than the spatially corresponding information in the SAX slices. This shows that the PCCN is able to utilize the available data from multiple views and select the best trade-off between the available information as the final output.

\section{Discussion}
\label{sec:discussion}

We have developed and successfully validated a fully automatic 4-step pipeline for cardiac surface reconstruction from raw cine MR images.
The PCCN as the main step of the pipeline is able to solve both the sparsity and misalignment issues in a single model, while retaining both class-specific information of the different cardiac substructures and cardiac phases (ED or ES). Its architecture is specifically designed for direct and effective point cloud processing. On the one hand, this enables a more memory-efficient data storage and the usage of higher resolutions to represent anatomical surfaces, which is beneficial for many downstream tasks \citep{beetz2021gen,beetz2021predicting,acero2022understanding,pamela2022characterizing}. On the other hand, the fact that only the surface level information is processed by the network facilitates the reconstruction task and ultimately leads to better performance than inefficient grid-based CNNs which require considerably larger amounts of memory to store the same 3D surface data and force the network to manage the additional difficulty of processing highly sparse data. No post-processing step needs to be applied on our reconstructed point clouds making its application easier than voxel grid-based approaches, which often require further processing (e.g. selection of largest connected component) \citep{xu2019ventricle}. While we develop the PCCN for three classes and two cardiac phases in this work, the network design can easily be extended to additional cardiac substructures or cardiac phases.

The point cloud-based deep learning approach also allows a straight-forward and effective integration of both SAX and LAX information which is crucial for an accurate 3D surface reconstruction, especially in information-sparse regions between slices or in the apical and basal areas of the heart. This in turn is of high importance for many downstream tasks, such as the accurate measurement of longitudinal strain which would be considerably more noisy when based only on SAX information. In addition, the PCCN can also be applied over manually delineated contours through a graphical interface \citep{banerjee2021ptrsa} providing flexibility to the first step of the reconstruction pipeline. Furthermore, the PCCN does not require any landmark detection, point-to-point correspondence or registration between the input and output point clouds for training, does not need any specific normalization to be applied to the input point clouds, and also does not rely on any template shapes, as opposed to many deformation-based reconstruction approaches (e.g. \citet{lamata2014automatic}).

Our PCCN achieves mean Chamfer distances between the reconstructed and gold standard point clouds that are below or similar as the underlying image resolution for all tested misalignment levels, cardiac substructures, and cardiac phases. This demonstrates that the PCCN is able to reconstruct a large variety of cardiac shapes that differ both spatially and temporally with high accuracy on both a local and global level. This is facilitated by the design of the PCCN decoder with both a coarse and dense output point cloud attending to information at different scales. The low standard deviation values of the Chamfer distances show that this high reconstruction quality is consistently obtained throughout the dataset indicating a high robustness of the network against outlier cases. Once trained, the PCCN also offers considerable speed advantages compared to traditional non deep learning-based reconstruction techniques \citep{lamata2014automatic,villard2018surface,banerjee2021ptrsa}, making it particularly advantageous for large-scale data processing. The combined multi-class anatomy processing is especially beneficial in this regard as it avoids the need for separate reconstruction processes to be run for each cardiac substructure.

We observe that the PCCN pre-trained on the 3D MRI-based SSM dataset can be successfully applied to the UKB dataset in a cross-domain transfer setting as part of the full reconstruction pipeline. This indicates that both the shape deformations and virtual slice planes selected during the creation of the SSM dataset are a realistic representation of real-world conditions. Furthermore, we did not observe any major negative bias or smoothing effects in the reconstructed shapes which showcases the suitability of the PCCN for cross-domain applications. As expected, we observe larger reconstruction errors for larger amounts of introduced misalignment which reflects the more difficult task. Male hearts generally show larger Chamfer distances than female ones across all misalignment levels and substructures. We believe this to be primarily a consequence of using the same point cloud resolution to represent the larger male hearts. This results in typically larger spatial distances between individual points even in case of similarly high reconstruction quality which is no longer present once the values are normalized by heart size. We find that the PCCN pre-trained on mildly misaligned SSM data achieves the best performance on the UKB dataset. This is somewhat surprising as the medium level was originally selected to reflect the average misalignment of typical acquisitions as in the UKB study. We hypothesize that on the one hand, the UK Biobank cohort could suffer from smaller amounts of misalignment than comparable studies due its usage of a coherent acquisition protocol or the selection of relatively healthy volunteers. On the other hand, the small misalignment amounts in the SSM dataset might also act as a regularizer during network training, which in turn helps the PCCN's generalization ability to the new UKB domain. It should also be noted that stronger misalignments are still present in the mildly misaligned SSM dataset albeit to a lesser extent. Finally, the selected misalignment amounts for each level are only chosen as an approximation derived from literature and could therefore also exhibit some degree of error. However, since the Chamfer distances show high reconstruction accuracy for all four PCCNs pre-trained on different misalignment levels, we conclude that a different choice in pre-training dataset would result in only a marginal performance drop. Furthermore, since there are no ground truth shapes available for the UK Biobank dataset, we base our evaluation on a comparison with a pseudo ground truth created by artificially introducing misalignment to selected real data. This likely results in a certain amount of noise in the pseudo ground truth and hence limits the accuracy of the obtained results. However, the misalignment was introduced in a way to approximate real conditions as closely as possible based on findings in prior work. Furthermore, we have also qualitatively assessed the quality of the artificial misalignment with a comparison to the true misalignment in the UK Biobank cases to ensure a high degree of realism. While the manual creation of a potentially more accurate ground truth is a possibility, this would also introduce a degree of subjectivity into the gold standard and significantly complicate the application to larger datasets.

Using the mild misalignment PCCN, we observe a high degree of alignment between the clinical metrics calculated directly from our 3D reconstructed meshes and the respective benchmark methods. This shows that both cardiac anatomy and function are accurately represented in the reconstructions while successfully taking into account the differences in subpopulations (sex, age), cardiac structures, and phases on a real-world dataset. It also further corroborates the accuracy of our pre-training and cross-domain transfer steps. Furthermore, it provides evidence of the effectiveness of our proposed meshing procedure, as a topologically correct two-manifold mesh is required for accurately calculating volumetric biomarkers. While no such topological correctness was achieved for some cases with the current approach, additional fine tuning of the relevant hyperparameters and pipeline would likely further improve the quality of the resulting meshes. 

The most noticeable differences in clinical metrics between the 3D and 2D-based calculations are found in the larger values obtained for the LV myocardial mass and RV volumetric metrics. The latter is an expected outcome that we believe to be a consequence of the general RV mesh shape in the original SSM by \citet{bai2015bi} that we used to derive our SSM dataset and pre-train our PCCN. In the SSM, the RV extends considerably above the basal SAX plane which leads to higher 3D volumes compared to a 2D-based calculation where the disk around the basal plane position serves as the boundary for calculating the respective volumes. These larger RV volumes are reflected in the reported scores in Tables~\ref{tab:clinical_metrics_comp_per_sex} and \ref{tab:clinical_metrics_female_age_groups}. This explanation is further corroborated by the RVEF values which show high similarity with the 2D-based approaches due to it being a relative metric that normalizes out raw size differences in volumes. We also note that the three comparative benchmarks rely on manual \citep{petersen2017reference}, semi-automatic \citep{bai2015bi}, and fully automatic (Simpson's rule applied to our UK Biobank dataset) approaches respectively to obtain the image segmentations required for their biomarker calculations. This further corroborates the good performance of our method, as its reconstructions exhibit similar clinical metrics as multiple ground truth benchmarks derived in different ways.

Finally, we find that the PCCN is able to successfully correct common errors in the segmentation contours of the precursor task by providing continuous and smooth myocardium boundaries with appropriate thickness even in cases of myocardial breakage or erroneous inclusion of the papillary muscles. This indicates that the PCCN is capable of implicitly learning an accurate anatomical prior during training which in turn allows it to automatically adjust anatomical inconsistencies.

\section{Conclusion}
\label{sec:conclusion}

We have developed a novel multi-class Point Cloud Completion Network capable of reconstructing 3D biventricular surface anatomies from sparse and misaligned cine MRI contours with high accuracy, while taking both temporal and spatial differences in the underlying cardiac substructures into account. We have also shown that the PCCN trained on a synthetic 3D MRI-based dataset can be successfully applied as the key component of a multi-step 3D cardiac surface reconstruction pipeline from raw 2D cine MRI acquisitions of the UK Biobank dataset in a cross-domain transfer setting. Finally, we have thoroughly evaluated both the PCCN and the complete 4-step pipeline on two different datasets and found very high reconstruction accuracy and robustness in terms of a variety of both geometric and clinical metrics. In our future works, we plan to investigate the possibility for further architectural improvements, for example by using the point cloud-based attention mechanisms, and to extend the presented method to other cardiac substructures and the full cardiac cycle. We also plan to evaluate the cardiac reconstruction performance over varying cardiac pathologies in the near future.

\section*{Acknowledgments}
This research has been conducted using the UK Biobank Resource under Application Number ‘40161’. The authors express no conflict of interest. The work of M. Beetz was supported by the Stiftung der Deutschen Wirtschaft (Foundation of German Business). A. Banerjee is a Royal Society University Research Fellow and is supported by the Royal Society Grant No. URF{\textbackslash}R1{\textbackslash}221314. The work of A. Banerjee and V. Grau was supported by the British Heart Foundation (BHF) Project under Grant PG/20/21/35082. The work of V. Grau was supported by the CompBioMed 2 Centre of Excellence in Computational Biomedicine (European Commission Horizon 2020 research and innovation programme, grant agreement No. 823712). The work of J. Ossenberg-Engels was supported by the Engineering and Physical Sciences Research Council (EPSRC) and Medical Research Council (MRC) [grant number EP/L016052/1]. The authors would like to thank Dr Wenjia Bai (Imperial College London) for providing the statistical shape model of the cardiac phases.

\appendix

\section{Additional implementation details}

\begin{table*}[!ht]
    \caption{\label{tab:clinical_metrics_male_age}Clinical metrics of male cases split by age group and calculated using different methods.}
    \def\arraystretch{1.5}\tabcolsep=3pt
        \centering
        \resizebox{\textwidth}{!}{
        \begin{tabular}{|p{55pt}|p{48pt}|p{42pt}|p{42pt}|p{48pt}|p{42pt}|p{42pt}|p{48pt}|p{42pt}|p{42pt}|}
            \hline
            {}      & \multicolumn{9}{c|}{Age groups (years)}  \\
            \cline{2-10}
            {}      & \multicolumn{3}{c|}{45-54}  & \multicolumn{3}{c|}{55-64} & \multicolumn{3}{c|}{65-74}  \\
			\cline{2-10}
            {}      & \citet{petersen2017reference} & Simpson's rule    & Proposed  & \citet{petersen2017reference}  & Simpson's rule    & Proposed & \citet{petersen2017reference}  & Simpson's rule    & Proposed\\
			\hline
			LVEDV (ml)      & 170   & 171 $\pm$ 27      & 170 $\pm$ 34      & 169   & 157 $\pm$ 31      & 158 $\pm$ 31      & 156   & 150 $\pm$ 28      & 147 $\pm$ 29  \\
			\hline
			LVESV (ml)      & 71    & 73 $\pm$ 14       & 74 $\pm$ 15       & 71    & 68 $\pm$ 18      & 67 $\pm$ 14      & 66   & 64 $\pm$ 20      & 64 $\pm$ 16  \\
			\hline
			LVSV (ml)       & 99    & 98 $\pm$ 17      & 96 $\pm$ 28       & 98    & 90 $\pm$ 18      & 91 $\pm$ 24      & 90   & 86 $\pm$ 16      & 83 $\pm$ 24  \\
			\hline
			LV mass (g)     & 106   & 104 $\pm$ 17      & 125 $\pm$ 24       & 104   & 100 $\pm$ 18      & 117 $\pm$ 24      & 99   & 97 $\pm$ 16      & 115 $\pm$ 21  \\
			\hline
			LVEF (\%)       & 58    & 57 $\pm$ 5        & 56 $\pm$ 8        & 58    & 57 $\pm$ 5      & 57 $\pm$ 7      & 58   & 58 $\pm$ 7      & 56 $\pm$ 9  \\
			\hline
			RVEDV (ml)      & 192   & 183 $\pm$ 30      & 208 $\pm$ 34      & 181   & 169 $\pm$ 34      & 194 $\pm$ 33      & 173   & 162 $\pm$ 29      & 185 $\pm$ 28  \\
			\hline
    		RVESV (ml)      & 91    & 84 $\pm$ 18      & 101 $\pm$ 16       & 82    & 76 $\pm$ 20      & 92 $\pm$ 18      & 81   & 73 $\pm$ 17      & 89 $\pm$ 18  \\
			\hline
			RVSV (ml)       & 101    & 99 $\pm$ 16      & 107 $\pm$ 25       & 98    & 92 $\pm$ 19      & 102 $\pm$ 22      & 92   & 89 $\pm$ 18      & 95 $\pm$ 20  \\
			\hline
			RVEF (\%)       & 53    & 54 $\pm$ 5      & 51 $\pm$ 6        & 55    & 55 $\pm$ 6      & 53 $\pm$ 6        & 54   & 55 $\pm$ 6      & 53 $\pm$ 7  \\
			\hline
            \multicolumn{5}{@{}l}{Values represent \emph{mean $\pm$ standard deviation}.}
    \end{tabular}
    }
\vspace*{-\baselineskip}
\end{table*}

\begin{table*}[!ht]
    \caption{Chamfer distances between dense ground truth point clouds and corresponding sparse and misaligned input contours in different SSM datasets with increasing levels of slice misalignment.}
 	\def\arraystretch{1.5}\tabcolsep=3pt
            \centering
 		\begin{tabular}{|p{100pt}|p{80pt}|p{80pt}|p{80pt}|p{80pt}|}
 			\hline
 			{} &  \multicolumn{4}{c|}{Misalignment Level}\\
 			\cline{2-5}
 			{}   & Mild & Medium & Strong & Severe\\
 			\hline
 			ED LV endo (mm) & 2.46 $\pm$ 0.26 & 3.27 $\pm$ 0.36 & 4.13 $\pm$ 0.54 & 5.22 $\pm$ 0.67 \\
			\hline
 			ED LV epi (mm) & 2.64 $\pm$ 0.29 & 3.43 $\pm$ 0.37 & 4.29 $\pm$ 0.53 & 5.38 $\pm$ 0.66 \\
 			\hline
 			ED RV (mm) & 3.60 $\pm$ 0.56 & 4.38 $\pm$ 0.60 & 5.26 $\pm$ 0.72 & 6.24 $\pm$ 0.82 \\
			\hline
 			ES LV endo (mm) & 2.45 $\pm$ 0.26 & 3.17 $\pm$ 0.37 & 3.95 $\pm$ 0.51 & 4.96 $\pm$ 0.74 \\
 			\hline
 			ES LV epi (mm) & 2.79 $\pm$ 0.29 & 3.49 $\pm$ 0.37 & 4.26 $\pm$ 0.50 & 5.28 $\pm$ 0.69 \\
 			\hline
 			ES RV (mm) & 3.25 $\pm$ 0.67 & 3.95 $\pm$ 0.64 & 4.72 $\pm$ 0.82 & 5.65 $\pm$ 0.93 \\
 			\hline
 			\multicolumn{5}{p{245pt}}{Values represent \textit{mean $\pm$ standard deviation}.}
 	      \end{tabular}
\label{tab:misalignment_levels_baseline_ssm}
\vspace*{-0.5\baselineskip}
\end{table*}

\subsection{Details of segmentation method for SAX stack and 4-chamber LAX}

The segmentation method proposed by \citet{bai2018automated} utilises a fully convolutional network (FCN) \citep{long2015fully}, where the neural network architecture learns image features from fine to coarse scales by applying a number of convolutional filters and combines the multi-scale features for predicting the class label at each image pixel. The network is adapted from the VGG-16 network \citep{simonyan2015very}, where each convolution uses a $3 \times 3$ kernel, followed by batch normalisation \citep{IoffeICML2015} and rectified linear unit (ReLU). After every two or three convolutions, the feature map is downsampled by a factor of $2$ in order to learn features at a more global scale. Feature maps learnt at different scales are upsampled to the original resolution using transposed convolutions and the multi-scale feature maps are then concatenated. Finally, three convolutional layers of kernel size $1 \times 1$, followed by a softmax function, are used to predict a probabilistic label map. The segmentation is determined at each pixel by the class label with highest softmax probability. The mean cross entropy between the probabilistic label map and the manually annotated label map is used as the loss function. We have used two pre-trained networks for segmenting the SAX slices and the 4-chamber LAX slices in our pipeline \citep{banerjee2021ptrsa}. Both networks were originally trained over more than $3750$ subjects from the UK Biobank study \citep{petersen2015uk}, with manual annotations of LV endocardial and epicardial borders and the RV endocardial borders at end-diastolic and end-systolic time frames. Data augmentation was performed on-the-fly, which applied random translation, rotation, scaling, and intensity variation to each mini-batch of 20 image slices, before feeding them to the network. The Adam method \citep{kingma2015adam} was used for optimising the loss function, with a learning rate of $0.001$ and iteration number of $50,000$.

\subsection{Details of conversion of 2D contours to 3D point cloud}
The septal wall for both SAX and 4ch LAX slices is identified as the intersection between LV epicardium and RV endocardium \citep{banerjee2021ptrsa}. As the SSMs described in Sec.~2.1 in the main manuscript do not contain the mitral valve, we identify it as part of the LV endocardial contour near the basal SAX plane and disconnected from the LV myocardium and remove it from both 4ch LAX and 2ch LAX slices for our reconstruction purpose. The basal SAX slice often contains both ventricular and atrial structures and so the pre-trained network sometimes provides suboptimal segmentation performance for the basal SAX slice. Hence, as a quality control, in case the LV myocardial region on the basal SAX slice is very small (less than 12 pixels) or the distance between the LV epicardial contour and RV endocardial contour is very high (more than 15~mm), we consider it to be an erroneous segmentation and remove the extracted contours from this slice for subsequent analyses.

\subsection{Details of surface mesh generation from dense point cloud}
We conduct various checks on the topology of the resulting meshes to ensure that each mesh contains the correct number of holes (one hole at the base of the LV endocardium and LV epicardium, no hole in the RV), consists of only one connected component, and fulfills all conditions required for a two-manifold mesh. In case of failed checks, we design an automated pipeline that dynamically adjusts the hyperparameter settings, re-executes the meshing step, and rechecks the topology. This process is repeated until a topologically correct two-manifold mesh has been created, ensuring a high quality mesh as the final output of the reconstruction pipeline.

\section{Relevant details of prior works on misalignment artifact}
\label{sup_sec:misalignment_prior_work}

In this section, we provide the numerical misalignment values reported in multiple pertinent prior works \citep{mcleish2002study,shechter2004respiratory,chandler2008correction,villard2016correction,xu2019ventricle,tarroni2020large} on motion-based slice misalignment in cardiovascular imaging and briefly discuss relevant differences with the setup in this work. 
The key statements regarding misalignment values in the referenced sources are as follows:
\begin{itemize}
\item \cite{xu2019ventricle}: ``To mimic the misalignment caused by motion artifacts, we kept fixed image planes and applied 3D rigid transformations to the model (random rotations no larger than 10° and random translations of no more than 4~mm) […].''
\item \cite{tarroni2020large}: ``Inter-slice misalignment (Fig.~5, left) had a median value of 2.29~mm and an interquantile range (IQR) of 1.17~mm.''
\item \cite{chang2020automatic}: ``The ranges of the translation in the x, y and z directions were $\pm 1.9$~mm, $\pm 3.6$~mm, and $\pm 12.2$~mm respectively, and the ranges of the rotations in the x, y and z directions were $\pm 0.8$~degrees, $\pm 3.2$~degrees, and $\pm 0.4$~degrees respectively.''
\item \cite{shechter2004respiratory} (based on Free Breathing Angiogram): ``For all patients, the heart translated caudally (mean, $4.9 \pm 1.9$~mm; range, 2.4 to 8.0~mm) and underwent a cranio-dorsal rotation (mean, $1.5\degree \pm 0.9\degree$; range, $0.2\degree$ to $3.5\degree$) during inspiration. In eight patients, the heart also translated anteriorly (mean, $1.3 \pm 1.8$~mm; range, -0.4 to 5.1 mm) and rotated in a caudo-dextral direction (mean, $1.2\degree \pm 1.3\degree$; range, $-1.9\degree$ to $3.2\degree$).''
\item \cite{mcleish2002study} (based on difference between maximum inhale and maximum exhale): ``[…] typical deformations were 3-4~mm with deformations of up to 7~mm observed in some subjects.''
\item \cite{villard2016correction}: ``Table 1 shows the mean, median, and standard deviation resulting from contour to contour distance calculations before the alignment […]: Median = 2.19; Mean = 2.82; Std = 2.48''.
\end{itemize}

\cite{shechter2004respiratory} reported values for respiratory motion artifact in free breathing angiograms. This is different from the cine MRI acquisitions with breath holds used in this work. \cite{mcleish2002study} measured motion values between maximum exhale and inhale positions: “The Volunteer Results (V) Show the Movement Between Maximum Exhale and Maximum Inhale”, “The patients were asked to hold their breath at the normal end-expiratory and the normal end-inspiratory positions”. Consequently, the reported values are likely considerably larger than what would typically be expected for a standard cine MRI acquisition with breath hold.

\section{Analysis of differences between 3D and 2D-based metrics calculations}

\begin{figure}[!t]
		\centerline{\includegraphics[width=0.485\textwidth]{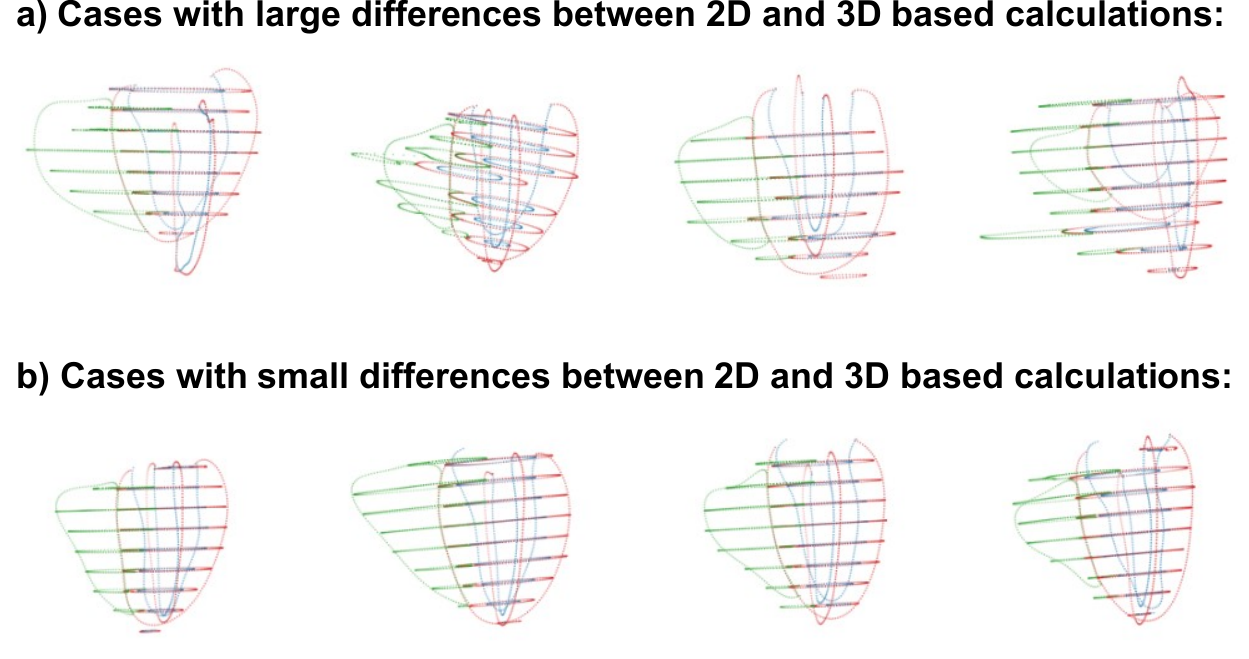}}
	\caption{Sample cases at ES with largest (a) and smallest (b) differences between 2D and 3D-based calculations of clinical metrics.}
	\label{fig:2d_vs_3d_metrics_calculation_differences}
 \vspace*{-0.4\baselineskip}
\end{figure}

In order to analyze the observed LV myocardial mass differences between the 3D and 2D-based calculations in greater detail, we select four UK Biobank cases with particularly large and small difference values and visualize them in Fig.~\ref{fig:2d_vs_3d_metrics_calculation_differences}-a and Fig.~\ref{fig:2d_vs_3d_metrics_calculation_differences}-b, respectively.
We find that LV mass differences are especially large in cases with high amounts of misalignment and low for cases with very little slice misalignment. 

We presume that these results are caused by the usage of misalignment correction in the calculations based on 3D reconstructions which is missing in the 2D-based approach and therefore leads to particularly large differences when stronger misalignment is present. 

\bibliographystyle{model2-names.bst}\biboptions{authoryear}
\bibliography{refs}

\end{document}

%% file: abstract-authors.tex
\author[1]{Marcel Beetz\corref{cor1}}
\ead{marcel.beetz@eng.ox.ac.uk}
\cortext[cor1]{Corresponding author}
\author[1,2]{Abhirup Banerjee}
\ead{abhirup.banerjee@eng.ox.ac.uk}
\author[1]{Julius Ossenberg-Engels}
\author[1]{Vicente Grau}

\address[1]{Institute of Biomedical Engineering, Department of Engineering Science, University of Oxford, Oxford OX3 7DQ, UK}
\address[2]{Division of Cardiovascular Medicine, Radcliffe Department of Medicine, University of Oxford, Oxford OX3 9DU, UK}

\begin{abstract}
Cine magnetic resonance imaging (MRI) is the current gold standard for the assessment of cardiac anatomy and function. However, it typically only acquires a set of two-dimensional (2D) slices of the underlying three-dimensional (3D) anatomy of the heart, thus limiting the understanding and analysis of both healthy and pathological cardiac morphology and physiology. In this paper, we propose a novel fully automatic surface reconstruction pipeline capable of reconstructing multi-class 3D cardiac anatomy meshes from raw cine MRI acquisitions. Its key component is a multi-class point cloud completion network (PCCN) capable of correcting both the sparsity and misalignment issues of the 3D reconstruction task in a unified model. We first evaluate the PCCN on a large synthetic dataset of biventricular anatomies and observe Chamfer distances between reconstructed and gold standard anatomies below or similar to the underlying image resolution for multiple levels of slice misalignment. Furthermore, we find a reduction in reconstruction error compared to a benchmark 3D U-Net by 32\% and 24\% in terms of Hausdorff distance and mean surface distance, respectively. We then apply the PCCN as part of our automated reconstruction pipeline to 1000 subjects from the UK Biobank study in a cross-domain transfer setting and demonstrate its ability to reconstruct accurate and topologically plausible biventricular heart meshes with clinical metrics comparable to the previous literature. Finally, we investigate the robustness of our proposed approach and observe its capacity to successfully handle multiple common outlier conditions.
\end{abstract}

\begin{keyword}
Cardiac 3D Surface Reconstruction \sep Multi-Class Point Cloud Completion Network \sep Cine MRI \sep Cross-Domain Transfer \sep Misalignment Correction \sep Geometric Deep Learning \sep Contours to Mesh Reconstruction

\end{keyword}